\documentclass[trackchanges,twocolumn]{aastex7}

\newcommand{\pla}{HAT-P-70b}
\usepackage{amsmath}

\begin{document}

\title{HAT-P-70b through the Eyes of MAROON-X: Constraining Elemental Abundances of Metals and Insights on Atmosphere Dynamics}

\correspondingauthor{Shi Lin Sun}

\author[0009-0004-6516-1873]{Shi Lin Sun}
\affiliation{Department of Physics and Trottier Institute for Research on Exoplanets, Universit\'{e} de Montr\'{e}al, 1375 Avenue Thérèse-Lavoie-Roux, Montreal, H2V 0B3, Canada}
\email[show]{shi.lin.sun@umontreal.ca}

\author[0000-0002-8573-805X]{Stefan Pelletier}
\affiliation{Observatoire astronomique de l'Université de Genève, 51 chemin Pegasi 1290 Versoix, Switzerland}
\email{stefan.pelletier@unige.ch}

\author[0000-0001-5578-1498]{Bj\"{o}rn Benneke}
\affiliation{Department of Earth, Planetary, and Space Sciences, University of California, Los Angeles, CA, USA}
\affiliation{Department of Physics and Trottier Institute for Research on Exoplanets, Universit\'{e} de Montr\'{e}al, 1375 Avenue Thérèse-Lavoie-Roux, Montreal, H2V 0B3, Canada}
\email{bbenneke@ucla.edu}

\author[0000-0001-7216-4846]{Bibiana Prinoth}
\affiliation{European Southern Observatory, Karl-Schwarzschild-Strasse 2, 85748 Garching bei München, Germany}
\affiliation{Lund Observatory, Division of Astrophysics, Department of Physics, Lund University, Box 118, 221 00 Lund, Sweden}
\email{bibiana.prinoth@eso.org}

\author[0000-0001-9521-6258]{Vivien Parmentier}
\affiliation{Laboratoire Lagrange, Observatoire de la Côte d’Azur, CNRS, Université Côte d’Azur, Nice, France}
\email{vivien.parmentier@oca.eu}

\author[orcid=0000-0003-4733-6532]{Jacob L.\ Bean}
\affiliation{Department of Astronomy and Astrophysics, University of Chicago, Chicago, IL 60637, USA}
\email{jacobbean@uchicago.edu}

\author[0000-0003-3191-2486]{Joost P. Wardenier}
\affiliation{Weltraumforschung und Planetologie, Physikalisches Institut, University of Bern, Gesellschaftsstrasse 6, 3012 Bern, Switzerland}
\affiliation{Department of Physics and Trottier Institute for Research on Exoplanets, Universit\'{e} de Montr\'{e}al, 1375 Avenue Thérèse-Lavoie-Roux, Montreal, H2V 0B3, Canada}
\email{joost.wardenier@unibe.ch}

\author[0000-0003-1728-8269]{Yayaati Chachan}
\affiliation{Department of Astronomy and Astrophysics, University of California, Santa Cruz, CA 95064, USA}
\email{ychachan@ucsc.edu}

\author[0000-0001-7329-3471]{Valentina Vaulato}
\affiliation{Observatoire astronomique de l'Université de Genève, 51 chemin Pegasi 1290 Versoix, Switzerland}
\email{valentina.vaulato@unige.ch}

\begin{abstract}
Ultra-hot Jupiters (UHJs) are exceptional laboratories for studying planetary atmospheres under extreme irradiation conditions. With close-in tidally locked orbits, these planets can have daysides hot enough for metals to be significantly ionized while still maintaining nightsides cold enough for refractory species to potentially condense. We present an analysis of the ultra-hot Jupiter HAT-P-70b taken with the MAROON-X high-resolution spectrograph. Using cross-correlations, we detect 14 neutral and singly ionized species, including Fe I, Fe II, Ti I, Ca I, Ca II, Cr I, Na I, V I, Mn I, Ni I, Mg I, Ba II, O I, and Sr I, with tentative evidence for H I, Co I, and K I. The absorption signals exhibit blueshifts on the order of a few $\mathrm{km\,s^{-1}}$, consistent with day-to-night winds. We further constrain relative abundances with atmospheric retrievals and demonstrate that some inferred elemental abundance ratios depend strongly on modeling assumptions. In particular, we show that a well-mixed retrieval approach neglecting ionization can strongly bias highly ionizable elements such as Ca and Ti. Accounting for the effects of equilibrium chemistry and thermal ionization generally results in inferred elemental abundance ratios that are closer to expectations for a solar-like composition, although not in all cases. Interestingly, we find a distinct nickel enrichment on HAT-P-70b, adding to the growing number of UHJ studies where the Ni abundance is seemingly enhanced. Our results underline the importance of considering physical and chemical atmospheric processes such as ionization when interpreting high-resolution transmission spectra of UHJs.

\end{abstract}


\section{Introduction}

Ultra-hot Jupiters (UHJs) are gas giants orbiting extremely close to their host stars. With orbital periods of a few days at most, they are tidally locked in synchronous rotation and have a hot, permanent dayside facing the star and a relatively cooler nightside facing away. Due to this proximity, their atmospheres are subject to strong stellar irradiation, leading to equilibrium temperatures ($\mathrm{T_{eq}}$) above $\sim2000\,\mathrm{K}$~\citep{Arcangeli2018, Bell2018, Parmentier2018}. 
At these high temperatures, refractory metals (Fe, Mg, Si, Ti, V, etc.) that would otherwise be condensed in cooler atmospheres can vaporize, adding significant optical opacity to the gas phase~\citep{Kitzmann2018}. 
The ionization of metals (Na, K, and Ca in particular) can also add significant amounts of free electrons to the atmosphere~\citep{savel_new_2024}, causing H$^-$ opacity (bound-free and free-free) to become important~\citep{Arcangeli2018, Vaulato2025}.
In tandem, important atmospheric coolants such as H$_2$O are significantly depleted from the upper dayside atmosphere due to thermal dissociation~\citep{Parmentier2018}.
As a result of added optical opacity sources under intense irradiation conditions and the removal of infrared coolants, UHJs can form strong thermal inversions~\citep[e.g.,][]{Hubeny2003, Fortney2008, Lothringer2018, gandhi_new_2019, piette_assessing_2020}.
Their hot, inflated atmosphere, where refractory elements are predominantly in the gas phase and thus spectroscopically accessible, makes UHJs ideal targets for probing the composition of giant planets, as well as the dynamics at play in these extreme atmospheres \citep{fortney_atmospheric_2021}.

High-resolution spectroscopy~\citep{Snellen2010, Brogi2012, Birkby2018, Snellen2025} has emerged as a powerful technique for characterizing the atmospheres of UHJs. It has a distinct advantage compared to current low or medium spectral resolving power instruments onboard space-based telescopes such as the HST and JWST. As opposed to molecules, most atomic and ionic species (e.g., Fe I, Fe II, Ca I, Ca II) only have very narrow absorption lines rather than broad-band features, which cannot be adequately probed using current space-based instrumentation. At resolving powers of $\mathrm{R} \ge 25000$, however, observations can detect individual spectral lines, resolve line-profiles, and measure wind-induced Doppler-shifts~\citep[e.g.,][]{Snellen2010, brogi_rotation_2016, Seidel2025, nortmann_crires_2025}. 
This has allowed high-resolution spectroscopy of ultra-hot Jupiters to reveal extremely rich observations of exoplanets comprising spectral signatures from a plethora of different chemical species simultaneously, giving unique insights into the thermal structures, dynamics, escape processes, and chemistry of their atmospheres~\citep[e.g.,][]{Hoeijmakers2019, Kesseli2022, Maguire2022}.

Among refractory metals routinely detected in UHJ atmospheres, two that are of particular interest are titanium and vanadium.  If present in the gas phase, the oxides of these species (TiO, and VO) have long been theorized to be potential drivers of thermal inversions in strongly irradiated giant exoplanets, owing to their strong opacities at optical wavelengths \citep{Hubeny2003, Fortney2008}. In practice, vanadium is consistently detected in UHJ atmospheres spanning a wide range of temperatures \citep{Prinoth2022, Borsato2023, Pelletier2023, Simonnin2025}. However, this is not the case for titanium for planets on the colder side of the UHJ population, potentially due to a cold trapping mechanism \citep{spiegel_can_2009, parmentier_3d_2013}. Previous studies find that on WASP-76b ($\mathrm{T_{eq}} \sim 2200\,\mathrm{K}$), Ti and TiO are not detected in cross-correlation analyses\citep{Tabernero2021, Kesseli2022}. In particular, retrieval results show that titanium is largely depleted , whether compared to Fe or equilibrium chemistry predictions \citep{Pelletier2023, Gandhi2023}. This suggests the presence of a titanium cold trap on the nightside of the planet, depleting titanium species from the observable gas phase. Similarly on WASP-121b ($\mathrm{T_{eq}} \sim 2350\,\mathrm{K}$), cross-correlation analyses did not find titanium signatures \citep{Merritt2020, hoeijmakers2024, Bazinet2025}. However, more recently, observations with ESPRESSO in 4UT mode reveal Ti I at high-significance on WASP-121b, but at a weaker level than predicted by models with respect to V I \citep{Prinoth2025}. A recent retrieval analysis on data from the Near Infrared Imager and Slitless Spectrograph on JWST shows that titanium is depleted by a factor of 10 or more, suggesting the presence of a partial titanium cold-trap that removes a fraction of titanium from WASP-121b's atmosphere~\citep{pelletier_enriched_2026}. 

In contrast, cross-correlation reveals strong signals from titanium species in the atmosphere of planets with higher equilibrium temperatures, such as TOI-1518b ($\mathrm{T_{eq}} \sim 2500\,\mathrm{K}$) \citep{Simonnin2025}, MASCARA-1b ($\mathrm{T_{eq}} \sim 2600\,\mathrm{K}$) \citep{Scandariato2023}, WASP-189b ($\mathrm{T_{eq}} \sim 2650\,\mathrm{K}$) \citep{Prinoth2022, Prinoth2023} and KELT-9b ($\mathrm{T_{eq}} \sim 4050\,\mathrm{K}$) \citep{Hoeijmakers2018, Zhang2026}. 
This trend suggests that the nightside cold trap mechanism that depletes titanium species from the observable atmosphere of cooler UHJs can be broken, at least partly, on hotter planets. However, the extent to which titanium is gradually released remains uncertain. Even on MASCARA-1b and WASP-189b, retrievals show that titanium is marginally underabundant \citep{Gandhi2023, Guo2024}. Crucially, the behavior of titanium relative to slightly less refractory but otherwise chemically similar species such as vanadium enable us to indirectly probe the condensation of nightside clouds, and the advection and vertical mixing processes capable of re-vaporizing them back to the observable atmosphere on the hotter dayside and terminator regions \citep{pelletier_enriched_2026}. 

With an equilibrium temperature of $2562\,\mathrm{K}$ \citep{Zhou2019}, \pla\ lies in a temperature regime where Ti is expected to present in gas form, although a fraction of its budget could still be missing if some regions of the nightside are still cold enough for it to condense. Early high-resolution cross-correlation spectroscopy (HRCCS) of a single transit obtained with the High Accuracy Radial velocity Planet Searcher for the Northern hemisphere (HARPS-N) spectrograph revealed a variety of neutral and ionized species, including V I \citep{BelloArufe2022}. While Ti I was not detected, a tentative signal of Ti II was reported, suggesting that \pla\ could be in the titanium transition region. Using the same dataset, \citet{Gandhi2023} performed a retrieval assuming vertically uniform abundance profiles and concluded that titanium is depleted in \pla's atmosphere compared to model predictions. More recently, thermal emission observations from CARMENES and PEPSI provided evidence for Ti I on the dayside of the planet via cross-correlation \citep{Guo2025}. Despite this detection, their retrievals recovered a subsolar Ti abundance, which was interpreted as evidence for the presence of a cold-trap removing some refractory elements from the gas phase. In parallel, time-resolved high-resolution spectroscopy with GHOST \citep{langeveld_time-resolved_2025} and CARMENES \citep{gan_revisiting_2026} identified strong Ca II absorption signals blueshifted by $\sim 5\mathrm{km\,s^{-1}}$, indicative of a partially ionized atmosphere with strong day-to-night winds at high altitudes. In this work, we revisit \pla\ with MAROON-X observations in the red-optical, with the goal of building a chemical inventory of the planet's atmosphere and constraining the abundances of metal species.

This paper is organized as follows. Section \ref{reduction} describes our observations and data reduction process. Section \ref{methods} details the cross-correlation and retrieval analyses. We present our results in Section \ref{results} and end with our conclusion in Section \ref{conclusions}.

\section{Observations and Data Reduction} \label{reduction}

HAT-P-70 is a fast-rotating ($v\mathrm{sin}i \sim 100\,\mathrm{km\,s^{-1}}$) A-type star with a magnitude of $\mathrm{V}=9.470 \pm 0.004$ and an effective temperature of $\mathrm{T_{eff}} = 8450^{+540}_{-690}\,\mathrm{K}$ \citep{Zhou2019}. \pla\ is an UHJ on a close-in orbit with a period of 2.74 days (Table \ref{system_params}). It is likely tidally-locked, with a hot, permanent dayside facing the star and a comparatively much cooler nightside facing away. Orbiting so close to its host star, \pla\ is subject to intense stellar irradiation, with an equilibrium temperature of $\mathrm{T_{eq} = 2562^{+43}_{-52}\,\mathrm{K}}$ assuming zero albedo and full heat redistribution \citep{Zhou2019}. The mass of \pla\ is poorly constrained from radial velocity measurements owing to its fast rotating host, with only an upper limit of $\mathrm{M_p} < 6.78\,\mathrm{M_{Jup}}$ at $3\sigma$ \citep{Zhou2019}.  More recently, \citet{Gandhi2023} spectroscopically derived a mass of $\mathrm{M_p} = 1.66^{+0.20}_{-0.20}\,\mathrm{M_{Jup}}$ from HARPS-N transit observations. \pla\ also has a large radius of $\mathrm{R_p} = 1.87^{+0.15}_{-0.10}\,\mathrm{R_{Jup}}$, likely indicating that its mass is no more than a few times that of Jupiter to hydrostatically support such an inflated radius~\citep{Gully-Santiago2024}. Here, $\mathrm{M_{Jup}}$ and $\mathrm{R_{Jup}}$ refer to one Jupiter mass and one Jupiter radius, respectively.

\begin{table}
\label{system_params}
\centering
\def\arraystretch{1.1}
\begin{tabular}{ccc}
\hline
\hline
Parameter & Value & Reference \\
\hline
\hline
Gaia RA (2015.5) & 4:48:12.560 & (1) \\
Gaia DEC (2015.5) & +09:59:52.726 & (1) \\
TYCHO V (mag) & $9.4700 \pm 0.0040$ & (1) \\
$M_*$ (M$_{\mathrm{\odot}}$) & $1.890^{+0.010}_{-0.013}$ & (1) \\
$R_*$ (R$_{\mathrm{\odot}}$) & $1.858^{+0.119}_{-0.091}$ & (1) \\
$T_\mathrm{eff}$ (K) & $8450^{+540}_{-690}$ & (1) \\
$\log g$ (cgs) & $4.181^{+0.055}_{-0.063}$ & (1) \\
$[$Fe$/$H$]$ & $-0.059^{+0.075}_{-0.088}$ & (1) \\
$v$sin$i$\,(km\,s$^{-1}$) & $99.85^{+0.64}_{-0.61}$ & (1) \\
$K_{*}$ (km\,s$^{-1}$) & $< 0.649\ (3\sigma)$ & (1) \\
$V_{\mathrm{sys}}$ (km\,s$^{-1}$) & $25.260 \pm 0.110$ & (1) \\
\hline
$P$ (days) &  $2.744320 \pm 0.000001$ & (2) \\
$T_0$\,(BJD$_\mathrm{TDB}$) & $2459175.05277 \pm 0.00017$ &  (2) \\
$M_p$ (M$_{\mathrm{Jup}}$) & $< 6.78\ (3\sigma)$ & (1) \\
$R_p$ (R$_{\mathrm{Jup}}$) & $1.87_{-0.10}^{+0.15}$  & (1) \\
$K_p$ (km\,s$^{-1}$) & $187$  & (3) \\
$T_\mathrm{eq}$ (K) & $2562^{+43}_{-52}$ & (1) \\
\hline
\hline
\end{tabular}\\ 
\caption{Star and planet properties for the HAT-P-70 system. References are as follows. (1) \citet{Zhou2019}, (2) \citet{ivshina_tess_2022}, (3) \citet{BelloArufe2022}} 
\end{table}

\begin{table*}
\label{observations}
\centering
\begin{tabular}{ccccccc}
\hline
\hline
Date  &  Duration [h]  &  \# of exposures &  $T_{\mathrm{exp}}$ [s]   &  Mean SNR  & Orbital phase range & \# of out-of-transit exposures\\
\hline
13 Dec 2023 & 4.6 & 36 & 403 (363) & 151 &  0.972 -- 0.041 & 9\\
24 Dec 2023 & 4.7 & 36 & 403 (363) & 133 &  0.970 -- 0.039 & 9\\
\hline
\hline
\end{tabular} \\
\caption{Overview of the MAROON-X observations of \pla. We indicate the values for the blue arm with the corresponding values for the red arm in parentheses. The reported mean SNR is per pixel.}
\end{table*}

We observed two transits of \pla\ using MAROON-X \citep{seifahrt_development_2016, seifahrt_maroon-x_2018, Seifahrt2020}, a high-resolution ($R \sim 85,000$), bench-mounted, fiber-fed échelle spectrograph operating in the red optical range ($500 - 920\ \mathrm{nm}$). MAROON-X is mounted on the $8.1\,\mathrm{m}$ Gemini North Telescope on Mauna Kea, Hawaii, and provides simultaneous coverage with two detectors: the ``blue” arm ($500 - 670\,\mathrm{nm}$) and the ``red” arm ($650 - 920\,\mathrm{nm}$). Observations were conducted on UT 2023-12-13 and UT 2023-12-24 (program: GN-2023B-Q-126, PI: Pelletier). For each night, we obtained 36 exposures, with the integration times of the blue and red arms set to $403\,\mathrm{s}$ and $363\,\mathrm{s}$, respectively, owing to their different readout times. The data were reduced using the standard MAROON-X reduction pipeline \citep{Seifahrt2020}, with the output organized as a data cube time series with dimensions $\mathrm{N_{exp} \times N_{order} \times N_{pixel}}$ for each transit and spectral channel, where $\mathrm{N_{exp}}$, $\mathrm{N_{order}}$ and $\mathrm{N_{pixel}}$ denote the number of exposures, spectral orders and detector pixel, respectively.

\begin{figure}[t]
  \centering
  \includegraphics[width=0.5\textwidth]{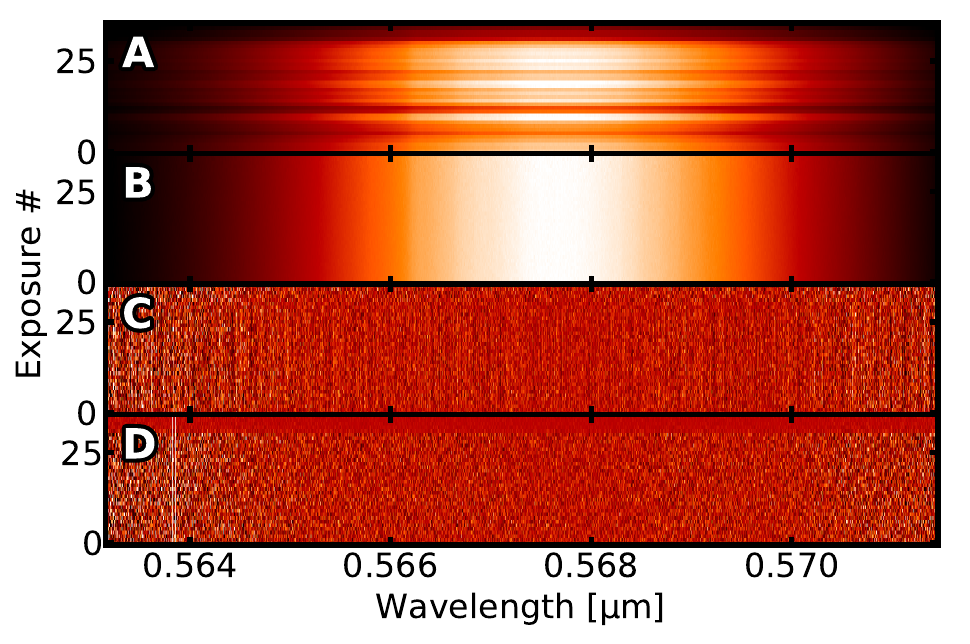}
  \caption{\textbf{Data detrending steps}. \textbf{A.} A single MAROON-X spectral order for the transit time series obtained on 24 December 2023. \textbf{B.} After outlier correction and continuum alignment. \textbf{C.} After dividing out the \textbf{polynomial fit to the} median out-of-transit. \textbf{D.} After PCA removal (first 5 components) of remaining stellar and telluric residuals. At this stage, the data contains the faint planetary trace as well as the Rossiter-McLaughlin (RM) trail buried in noise, and is served as input for the cross-correlation and retrieval analyses.}
  \label{detrending steps}
\end{figure}

\section{Methods} \label{methods}

In this section, we describe the data reduction and atmospheric modeling used to obtain a chemical inventory of \pla's atmosphere and constraints on the elemental abundances of species of interest.

\subsection{Data Detrending}

The observed MAROON-X spectroscopic time series contain the faint planetary signal mixed with the stellar spectrum, Earth's tellurics, and noise (Figure \ref{detrending steps}, panel A). Our goal is to remove the stellar and telluric contributions, leaving only the absorption signal of the planet buried in the noise. This is possible because the planetary signal undergoes radial velocity shifts of $\sim 60\,\mathrm{km\,s^{-1}}$ throughout the transit event, whereas the telluric and stellar lines remain stationary or quasi-stationary ($< 1\,\mathrm{km\,s^{-1}}$) in velocity space. Given the resolving power of MAROON-X ($R \sim 85,000$), which corresponds to a velocity resolution of $\sim 3.5\ \mathrm{km\ s^{-1}}$, the planetary Doppler shift is well captured by the instrument.

To remove unwanted telluric and stellar contributions, we follow a detrending procedure similar to \citet{Pelletier2023}. As the main source of contamination for atmospheric studies of metals in the atmosphere of \pla\ is expected to be the host star rather than the tellurics in the optical bandpass, we first shift the data into the stellar rest frame. We then mask out all low signal-to-noise ratio (SNR) wavelength regions where the flux drops to below 50\% that of the local continuum. In addition, we mask out wavelengths between $0.6868 - 0.694\,\mathrm{\mu m}$ and $0.7595 - 0.769\,\mathrm{\mu m}$, which correspond to the $\mathrm{O_2}$ B-band and A-band, respectively, from Earth's atmosphere \citep{Arnold2014, Bertaux2014}. 

To correct for outlier pixels, we divide each order of each exposure by its median flux. We then divide each spectral channel by its temporal median, effectively producing a residual matrix. In this residual matrix, outliers are flagged as pixels deviating by more than $6\sigma$ from the median of their corresponding spectral channel. These flagged pixels are subsequently corrected by interpolation from neighboring pixels along the spectral direction as in \citet{Brogi2014}. Finally, with all outlier pixels corrected, we multiply the residual matrix by the previously divided temporal and spectral medians to reproduce the now outlier corrected original flux matrix.

Following \citet{Gibson2020}, we then remove throughput and blaze variations by aligning all spectra to a common continuum level. More specifically, we first divide each pixel by its temporal median, then run a low-pass box plus Gaussian filter replicating continuum variations, and finally divide the original spectra by this filter. The box filter has a width of 51 pixels and is smoothed by a Gaussian kernel with a standard deviation of 100 pixels (Figure \ref{detrending steps}, panel B). We further compute the median spectrum of the out-of-transit exposures $M$, and divide out each spectrum by a second-order polynomial fit to $M$ of the form $aM^2 + bM + c$ (Figure \ref{detrending steps}, panel C).

Finally, as stellar and telluric lines are not perfectly removed by the median fit, we use a Principal Component Analysis (PCA) approach along the time axis to remove remaining correlated residuals. 
We find that while removing none or just a few principal components results in strong residuals remaining in the data, removing between 5 and 15 components yields similar and stable results. On the other hand, removing even more components risks suppressing the planetary signal. 
To avoid over-detrending, we choose to only remove the first five components in our analysis. We do so by taking the continuum-normalized spectrum and divide out each of the omitted components. At this stage, the data are mostly free of low acceleration stellar and telluric contamination while still carrying the high acceleration planetary trace and the Rossiter-McLaughlin (RM) trail, alongside intrinsic noise (Figure \ref{detrending steps}, panel D). The PCA notably cannot remove RM-induced residuals as they vary significantly in velocity throughout the transit, owing to the fast rotation rate of the host star~\citep{Zhou2019}.

The uncertainty for each pixel is estimated using a likelihood-minimization approach described in \citet{Gibson2020, gibson_relative_2022}, given by

\begin{equation}
\sigma_n = (aF_n + b)^{0.5},
\end{equation}
where $F_n$ is the measured flux at pixel $n$, and $a$ and $b$ are the fitted parameters. The fitting of $a$ and $b$ is done during the data detrending stage.

While the detrending procedure is effective at isolating the planetary signal from unwanted sources of contamination, it also distorts the data \citep{brogi_retrieving_2019}. Therefore, as in \citet{Pelletier2021}, all detrending steps are later reapplied on each forward model to mimic the distortions and avoid biases in the retrieved abundances.

\subsection{Atmosphere Modeling}

We use the SCARLET framework to generate atmosphere models of \pla, which serve as templates for our cross-correlation analysis \citep{Benneke2012, Benneke2013, Benneke2015, Benneke2019a, Benneke2019b, Pelletier2021, Pelletier2023, Bazinet2024}. We include the opacities of neutral and singly ionized species from the VALD database \citep{piskunov_vald:_1995, ryabchikova_major_2015}, as well as TiO \citep{McKemmish2019}, VO \citep{McKemmish2016}, and H$^-$ (bound-free and free-free) \citep{Gray2021}. Cross sections are calculated using HELIOS-K \citep{Grimm2015, Grimm2021}. We include in our atmosphere model the effects of collision-induced absorptions from $\mathrm{H_2} \textendash \mathrm{H_2}$ as well as $\mathrm{H_2} \textendash \mathrm{He}$ interactions \citep{Borysow2002}. All models are generated at a spectral resolving power of $\mathrm{R} = 250,000$ and then convolved with a Gaussian kernel to match the MAROON-X resolution of $\mathrm{R} = 85,000$.

\begin{figure}[t]
  \centering
  \includegraphics[width=0.5\textwidth]{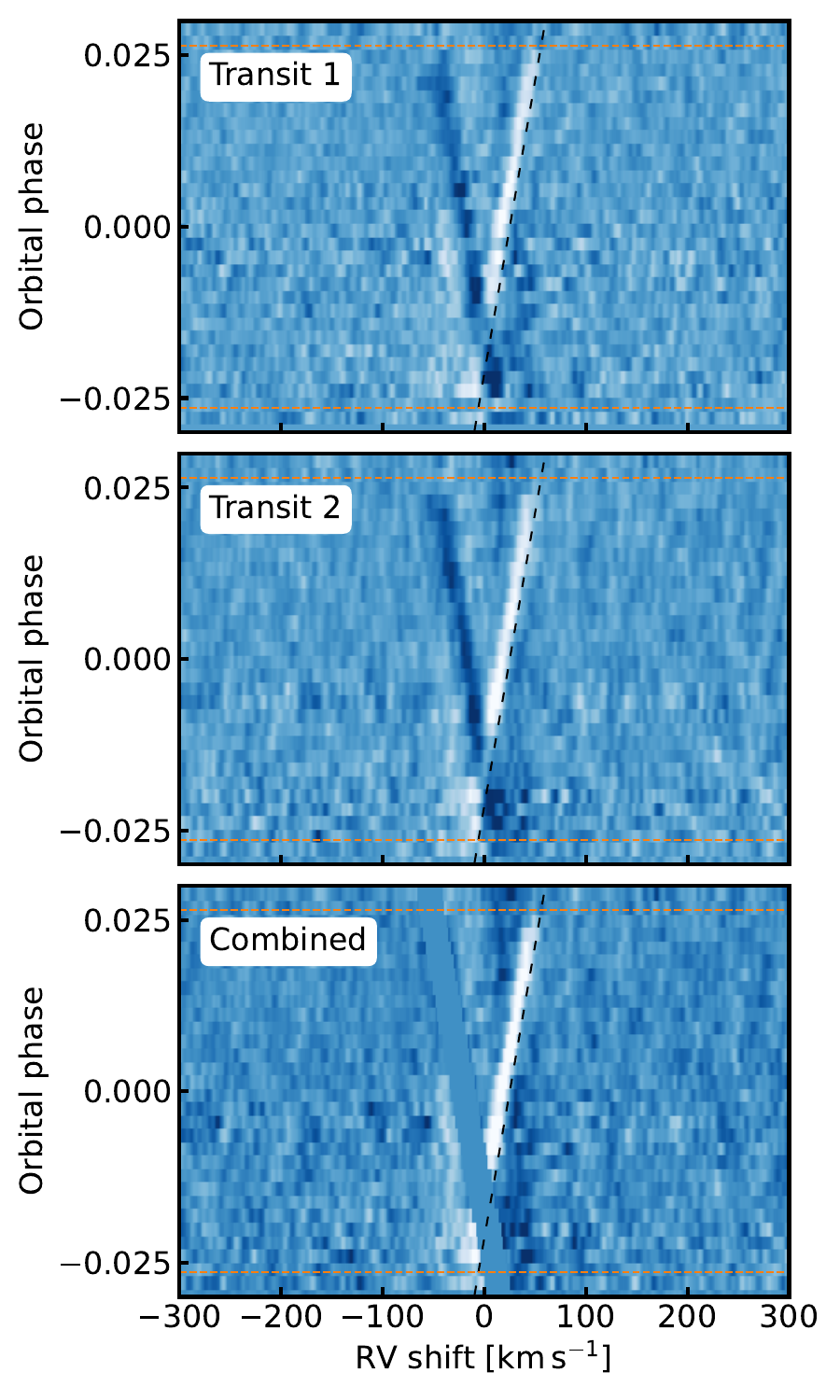}
  \caption{\textbf{Cross-Correlation time series of 2 transits of \pla\ from the MAROON-X spectrograph.} \textbf{Top panel:} CCF results of the night of 13 December 2023. The orange horizontal dashed lines mark the beginning and the end of the transit event. The slanted black dashed line shows the expected trail of the planet. The planetary signal shows as a white trace near the expected path. We see a clear dark trail, symmetric to the planetary signal with respect to zero velocity shift due to the RM effect.
  \textbf{Middle panel:} Same as top panel, but for the night of 24 December 2023. \textbf{Bottom panel:} Same as previous panels, but with both transits stacked and with the Rossiter-McLaughlin trail masked out.} 
  \label{RM mask}
\end{figure}

Ultra-hot Jupiter atmospheres are expected to be thermally inverted~\citep{Fortney2008, petz_pepsi_2025}, with the temperature increasing with altitude. However, the extent of the inversion is expected to strongly depend on the atmospheric composition, specifically the presence of optical absorbers (e.g., metal oxides and hydrides, neutral metals, H$^{-}$). To explore different temperature regimes, we follow a similar approach as in \cite{Prinoth2023} and generate three atmospheric models assuming isothermal temperature-pressure profiles of $2562\,\mathrm{K}$ (the equilibrium temperature of the planet), $4000\,\mathrm{K}$ and $6000\,\mathrm{K}$. The cooler atmosphere template is used to cross-correlate with the species expected to be more abundant in the lower atmosphere (e.g. neutral metals), while the hotter template at $4000\,\mathrm{K}$ is used to search for species expected to be more abundant in the upper atmosphere (e.g. ions). Finally, we include a template at $6000\,\mathrm{K}$ to cross-correlate with dissociation byproducts H and O that may be present in extended, potentially non-hydrostatic regions of the atmosphere. In particular, the 6000\,K templates of H and O show additional activated spectral features not present in the 4000\,K template, which could be present in the data and lead to a non-detection if not included.

For the three models used in the cross correlation analysis, we assume volume mixing ratios (VMRs) following chemical equilibrium predictions at the given temperatures for a solar-like composition, calculated with \texttt{FastChem} \citep{Stock2018, Stock2022, kitzmann_fastchem_2024}. 

\subsection{Cross-correlation Analysis}

We perform a cross-correlation analysis to search for species in the observed transmission spectrum of \pla. This technique has been widely used in the literature and has been proven to be highly effective in detecting a variety of metal species on UHJs \citep[e.g.,][]{Hoeijmakers2018, Hoeijmakers2019, Merritt2021}. The cross-correlation function (CCF) is given by

\begin{equation}
    \mathrm{CCF}(v) = \sum_n \frac{d_nm_n(v)}{\sigma_n^2},
\end{equation}
a summation over all wavelength bins (n), where $d_n$ is the detrended data (e.g.\ Figure~\ref{detrending steps}, panel D) with corresponding uncertainties $\sigma_n$, and $m_n(v)$ is the model transmission spectrum shifted by velocity $v$ \citep{Snellen2010}. We compute the CCF over a grid of velocity shifts ranging from $-400$ to $400\,\mathrm{km\,s^{-1}}$ in steps of $1\,\mathrm{km\,s^{-1}}$. 
While the data detrending is done order-by-order, the CCF is computed order by order, and then summed over all orders to produce a time-series computed over a range of velocities.

As our cross-correlation is performed with the data in the stellar reference frame, we take the extra step of aligning the CCF in the barycentric frame before combining both transits. This Barycentric Earth Radial Velocity (BERV) correction is done on the CCFs to avoid further interpolating the data directly.  Since many chemical species are expected to be present in both the star and the planet (e.g., Fe, Ca, Mg), we apply a mask with a width of $26\,\mathrm{km\,s^{-1}}$ along the visible dark RM trace to prevent contamination in our cross-correlation analyses (Figure \ref{RM mask}, Bottom panel). The width of the mask is chosen to preserve as much data as possible while removing the RM trace.  
Likelihood contributions from exposures where the planetary trace overlaps in velocity space with this mask are set to zero in the retrievals to avoid any biases.
We then phase-fold the data and sum the contributions across exposures over a grid of $K_p$ and $V_\mathrm{{sys}}$, which are the planetary radial velocity semi-amplitude and systemic velocity, respectively. The resulting phase-folded CCF in a $K_p\ \textendash\ V_{\mathrm{sys}}$ map should then peak near the expected velocity parameters of \pla\ ($K_p = 187 \pm 4\,\mathrm{km\,s^{-1}}$, \citealp{Gandhi2023}; $V_\mathrm{sys} = 21.76 \pm 0.39\,\mathrm{km\,s^{-1}}$, J. V. Seidel et al. Submitted) if the atmosphere models match the data. The signal-to-noise ratio (SNR) is computed as the CCF value at each position in the map divided by the estimated standard deviation away from the peak. We calculate the standard deviation from a box delimited by $V_\mathrm{sys} = -100\,\mathrm{km\,s^{-1}}$ and $V_\mathrm{sys} = 0\,\mathrm{km\,s^{-1}}$ to avoid the planetary and RM signals contributing to the noise estimate.

\subsection{Retrieval Analysis}

While a cross-correlation analysis is effective at confirming whether a species is present in \pla's atmosphere, it does not directly constrain abundances. Therefore, to quantify the atmospheric composition, we perform a retrieval analysis to constrain the elemental abundances of key species: Fe, Ti, Ca, Cr, Ni, Mg, V, Mn, Co and Sr. We also include H$^-$ as an opacity source in our retrievals, as H$^{-}$ bound-free absorption is expected to be the dominant source of continuum opacity at optical wavelengths in ultra-hot Jupiter atmospheres. We do not fit any cloud parameters as the dayside and terminator of \pla\ are likely too hot for any significant cloud mass to be sustainable~\citep{helling_cloud_2021}.

For our atmospheric retrievals, we use the SCARLET framework with \verb|emcee| as a sampler \citep{foreman-mackey_emcee_2013}. We follow the prescription of \cite{Gibson2020}, with the log-likelihood given by 

\begin{equation}
    \mathrm{ln}(L) = -\frac{N}{2} \mathrm{ln} \left[ \frac{1}{N} \left(\sum_n \frac{d_n^2 + m_n^2}{\sigma_n^2} - 2\mathrm{CCF} \right) \right],
\end{equation}
where $N$ is the total number of data points. To retrieve a temperature-pressure profile, we freely fit 8 evenly spaced points in log pressure between $10^{-10}$ and 1 bar, using the parameterization of \citet{Pelletier2021}. We choose a relatively weak smoothing prior of $\sigma_s = 500\,\mathrm{K\,dex^{-2}}$ to flexibly fit TP profiles with any shape while still penalizing nonphysically steep temperature gradient changes. We additionally retrieve the global $K_p$ and $V_\mathrm{sys}$, as well as the line broadening due to planetary rotation ($v_\mathrm{rot}$).

While many retrieval prescriptions exist in the literature for exoplanet atmosphere characterization, the most common ones fall under two main families: free (or well-mixed; \citealp[e.g.][]{barstow_outstanding_2020}) and chemical equilibrium \citep[e.g.][]{brogi_roasting_2023} retrievals. In this work, we explore both types of retrieval and showcase potential biases in inferred elemental abundances in both cases.

Because the mass of \pla\ is only reported as an upper limit ($\mathrm{M_p} < 6.78\,\mathrm{M_{Jup}}$; \citealp{Zhou2019}), one could include $\mathrm{M_p}$ as a free parameter in the retrieval. However, we note that the planetary mass is strongly degenerate with both the temperature structure and atmospheric abundances through the scale height. Variations of $\mathrm{M_p}$ can be largely compensated by adjustments in the temperature-pressure profile or species abundances, limiting the ability of high-resolution transmission spectroscopy to independently constrain the mass. Therefore, we do not treat it as a free parameter in the free retrieval and fix it at $\mathrm{M_p}$ = $1.66\,\mathrm{M_{Jup}}$, which corresponds to the spectroscopically derived value reported by \citet{Gandhi2023}.

In contrast, for the chemical equilibrium retrieval, we keep $\mathrm{M_p}$ as a free parameter. In this framework, the coupling between chemical species imposed by equilibrium chemistry can partially alleviate degeneracies, allowing the data to provide better constraints on the surface gravity. We use a uniform prior between $1\,\mathrm{M_{Jup}}$ and $6.64\,\mathrm{M_{Jup}}$, with an initial guess of $1.66\,\mathrm{M_{Jup}}$. The lower bound is chosen to ensure hydrostatic equilibrium at low pressures.

\begin{table}[t]
\centering
\renewcommand{\arraystretch}{1.1}
\setlength{\tabcolsep}{4pt}
\begin{tabular}{ccc}
\hline
\hline
Parameter & Description & Prior range\\
\hline
\hline
T$_i$ & Temperature of layer i [K] & $\mathcal{U}$[100, 8500]\\
$K_p$ & RV semi-amplitude [$\mathrm{km\,s^{-1}}$] & $\mathcal{U}$[157, 217]\\
$V_{\mathrm{sys}}$ & Systemic velocity [$\mathrm{km\,s^{-1}}$] & $\mathcal{U}$[-1.3, 38.7]\\
$v_\mathrm{rot}$ & Rotational broadening [$\mathrm{km\,s^{-1}}$] & $\mathcal{U}$[0, 20]\\
\hline
& Free &\\
\hline
logVMR$_j$ & VMR of species j & $\mathcal{U}$[-12, -1]\\
$\mathrm{M_p}$ & Planetary mass [$\mathrm{M_{Jup}}$] & $\mathcal{U}$[1.00, 6.64]\\
\hline
& Chemical equilibrium &\\
\hline
[M$_k$/H] & Metallicity of species k & $\mathcal{U}$[-3, 3]\\
\hline
\hline
\end{tabular}
\caption{\textbf{Retrieved parameter description and prior ranges for the free and chemical equilibrium retrievals.}
\newline
i = 0 to 7 (T$_0$ at 10$^{-10}$ bar and T$_7$ at 1 bar)
\newline
j = Fe, Ti, TiO, Ca, Cr, Ni, Mg, V, VO, Mn, Co, Sr, and H-
\newline
k = Fe-H$^-$, Ti-TiO, V-VO, Ca, Cr, Ni, Mg, Mn, Co, and Sr
}
\label{retrieval setup}
\end{table}

\subsubsection{Free retrieval}

The first prescription we explore is the classical free retrieval, which adopts a more agnostic approach. That is, the abundances of individual chemical species are assumed to be constant with altitude and treated as independent parameters, unconstrained by thermochemical equilibrium relations. This gives the retrieval flexibility to probe a wide range of possible atmospheric conditions that may include disequilibrium processes such as quenching and photochemistry. Although free retrievals often result in a larger parameter space, they provide a more data-driven measurement of the composition and allow deviations from equilibrium expectations.

In this work, we independently fit the log VMR of each species. This includes Fe, Ti, TiO, Ca, Cr, Ni, Mg, V, VO, Mn, Co, Sr and $\mathrm{H^-}$. In each layer of the atmosphere model, the retrieval sums the VMR of these species and adds $\mathrm{H_2}$, He and H as filler gases such that the VMR of all the species sum to one. As in \cite{Pelletier2023}, the relative amounts of $\mathrm{H_2}$, He and H are calculated using \texttt{FastChem} given the temperature-pressure profile queried at each step of the retrieval. 

Ionized species (e.g.: Fe II, Ca II) have very strong lines that probe into the extended upper atmosphere, a region that is likely not in hydrostatic equilibrium and which our models cannot match \citep{Maguire2022, deibert_exogems_2023, Prinoth2024, langeveld_time-resolved_2025, Simonnin2025}. As in \citet{Prinoth2023}, in the case of the similar UHJ WASP-189b in terms of equilibrium temperature, we find that our models cannot reproduce the strength of most ionic signals (e.g. Fe$^+$, Na$^+$). Indeed, we find that all test retrievals that include ions recover extremely high temperatures, which is likely more representative of the exosphere. As a result, we opt to fit only the opacity contributions of neutral species in this work.

In this framework, $\mathrm{M_p}$ is fixed, such that variations in the scale height are completely absorbed by the retrieved temperature structure and chemical abundances.

\subsubsection{Chemical Equilibrium retrieval} \label{chemequi description}

In contrast to free retrievals, chemical equilibrium retrievals link the abundances of all chemical species to a common set of atmospheric parameters (e.g., metallicity and C/O) to enforce thermochemical equilibrium given the queried temperature-pressure profile. This approach has the advantage of being more physically motivated while reducing the number of model parameters. In the context of UHJ atmospheres, the abundance profiles of many metal species are shaped by thermal dissociation, and chemical equilibrium retrievals have a natural way of incorporating this information. However, they lack the flexibility to model important disequilibrium processes, for example due to cold-trapping~\citep[][]{Pelletier2023} or quenching~\citep[][]{welbanks_high_2024, sing_warm_2024}.

The second retrieval prescription we test is a variation of the chemical equilibrium retrieval. Instead of having a global metallicity knob that controls the abundance of every species assuming solar-like proportions, we fit a set of elemental abundance, one for each metal species ([Fe/H], [Ca/H], [Ti/H], [V/H], [Mg/H], [Mn/H], [Cr/H], [Ni/H], [Co/H], and [Sr/H]), similar to \citet{Smith2024}. Here, for instance, [Fe/H] sets the abundance of neutral Fe while [Ti/H] controls both neutral Ti and TiO. The abundance profiles of all species are then calculated assuming equilibrium chemistry given these elemental abundances and the queried temperature-pressure profile. For consistency, both Fe and $\mathrm{H^-}$ are tied to the same elemental abundance parameter [Fe/H], as the $\mathrm{H^-}$ abundance is expected to depend on the availability of free electrons from ionization of metals. This effectively links the continuum opacity to the electron budget set by the metal content in \pla's atmosphere, while avoiding an additional free parameter in the retrieval. Similarly to the setup in our free retrieval, we only consider the opacity contributions of the neutral species in our models. However, we note that in this case, ionization is considered in the calculation of the abundance profiles. This approach grants more flexibility for the retrieval to model non-solar relative metal abundance while still naturally accounting for expected thermal ionization and dissociation processes~\citep[e.g.,][]{pelletier_crires_2025, pelletier_enriched_2026}.

\begin{figure*}[t]
  \centering
  \includegraphics[width=\textwidth]{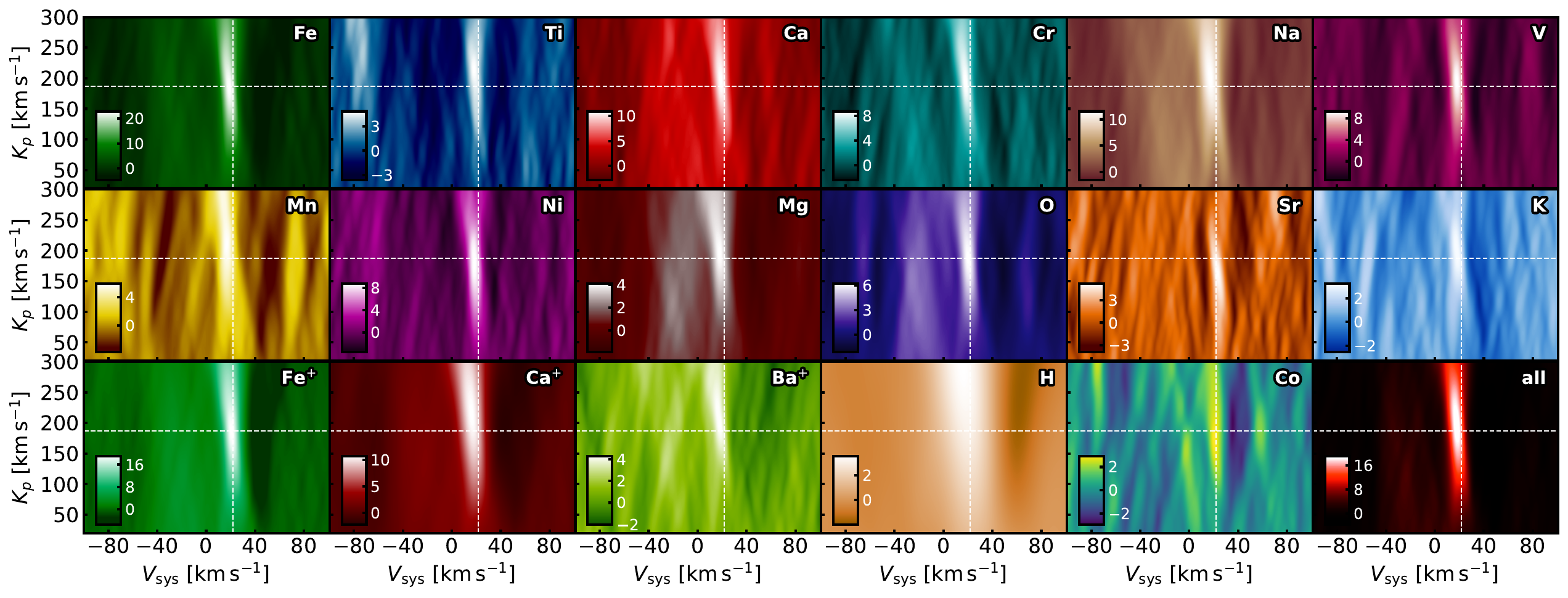}
  \caption{$\boldsymbol{K_p\ \textendash\ V_\mathrm{sys}}$\ \textbf{maps for species which we claim a detection or a tentative detection in \pla's atmosphere with 2 transits from the MAROON-X spectrograph.} White dashed lines indicate the expected position of the signal, assuming a symmetric planet with a static atmosphere. Peaks of the cross-correlation function are shown as white regions near the expected position. The peaks show a slight offset in $V_\mathrm{sys}$, indicative of the presence of strong winds in the atmosphere. The standard deviation is calculated with from a box delimited by $V_\mathrm{sys} = -100\ \mathrm{km\ s^{-1}}$ and $V_\mathrm{sys} = 0\ \mathrm{km\ s^{-1}}$}
  \label{detections}
\end{figure*}

\section{Results and Discussion} \label{results}

\begin{table}[t]
\centering
\renewcommand{\arraystretch}{1.1}
\setlength{\tabcolsep}{8pt}
\begin{tabular}{cccc}
\hline
\hline
Species & Max SNR & $K_p\,(\mathrm{km\,s^{-1}})$ & $\Delta V_\mathrm{sys}\,(\mathrm{km\,s^{-1}})$ \\
\hline
Fe I & $22.9$ & $185^{+21}_{-19}$ & $-3.5^{+1.4}_{-1.4}$ \\
Ti I & $4.9$ & $215^{+46}_{-51}$ & $-4.1^{+3.3}_{-2.8}$ \\
Ca I & $10.9$ & $198^{+39}_{-32}$ & $-3.3^{+2.1}_{-2.3}$ \\
Cr I & $8.7$ & $195^{+33}_{-30}$ & $-3.9^{+2.1}_{-2.1}$ \\
Na I & $11.5$ & $200^{+43}_{-34}$ & $-5.2^{+3.0}_{-3.1}$ \\
V I & $9.2$ & $195^{+30}_{-32}$ & $-4.3^{+2.1}_{-2.1}$ \\
Mn I & $5.9$ & $209^{+66}_{-49}$ & $-5.5^{+2.8}_{-2.8}$ \\
Ni I & $8.7$ & $179^{+41}_{-39}$ & $-3.7^{+2.1}_{-2.1}$ \\
Mg I & $4.1$ & $197^{+103}_{-63}$ & $-4.3^{+4.3}_{-4.4}$ \\
O I & $6.2$ & $188^{+51}_{-45}$ & $-1.7^{+2.9}_{-3.1}$ \\
Sr I & $5.2$ & $164^{+27}_{-31}$ & $0.8^{+2.2}_{-2.2}$ \\
Fe II & $19.0$ & $187^{+27}_{-23}$ & $-1.7^{+1.7}_{-1.6}$ \\
Ca II & $10.7$ & $218^{+61}_{-50}$ & $-6.3^{+3.8}_{-3.7}$ \\
Ba II & $4.3$ & $199^{+76}_{-45}$ & $-4.2^{+3.2}_{-3.9}$ \\
H I\ * & $3.5$ & $-$ & $-6.0^{+13.0}_{-7.1}$ \\
Co I\ * & $2.9$ & $175^{+127}_{-52}$ & $-1.6^{+4.1}_{-3.4}$ \\
K I\ * & $3.2$ & $189^{+63}_{-46}$ & $-3.8^{+3.3}_{-3.8}$ \\
All & $19.0$ & $196^{+28}_{-27}$ & $-4.3^{+1.7}_{-1.8}$ \\
\hline
\hline
\end{tabular}
\caption{\textbf{Cross-correlation analysis results for detected species.} We report the values of the maximum signal-to-noise ratio ($\mathrm{SNR_{max}}$) and the corresponding $K_p$ and $\Delta V_\mathrm{sys}$ (with respect to the stellar reference frame) for each species that we claim a detection or a tentative detection (indicated with an asterisk). We derive the uncertainties in $K_p$ and $\Delta V_\mathrm{sys}$ using the best-fit slices along both directions, with values corresponding to the positions at $\mathrm{SNR = SNR_{max} - 1}$. For H I, since we do not have significant constraints on $K_p$, we use instead the slice corresponding to the expected $K_p$ of the planet ($187\,\mathrm{km\,s^{-1}}$) to calculate the uncertainties in $\Delta V_\mathrm{sys}$. The majority of species are blueshifted relative to the systemic velocity of the star $V_\mathrm{sys} = 21.76\,\mathrm{km\,s^{-1}}$.} 
\label{CCF results}
\end{table}

\subsection{Cross-correlation Results}

Via a cross-correlation analysis, we search for every neutral and singly ionized element with available opacities in the VALD database, as well as TiO and VO, to produce a chemical inventory of species in \pla's atmosphere that are observable in the red optical. We confirm the detections of Ca II, Cr I, Fe I, Fe II, H I, Mg I, Na I and V I reported by \citet{BelloArufe2022} and further detect Ba II, Ca I, Mn I, Ni I, O I, Sr I, and Ti I (Figure \ref{detections}). We further find a tentative signal for Co I and K I, although the SNR is too low to claim a detection. We note that observations from CARMENES have also found a tentative signal of K I \citep{gan_revisiting_2026}. The detection of these neutral and ionized species is broadly consistent with previous results for other UHJs \citep{Kesseli2022, Borsato2023, Pelletier2023, Prinoth2025}. Notable non-detections are shown in the appendices (Figure \ref{non-detections}). However, these may simply due to the wavelength coverage of MAROON-X being limited to the red optical.
For example, while we find clear evidence of Ti I, we could not confirm the tentative Ti II signal observed from HARPS-N by \citet{BelloArufe2022}. This is because the strongest absorption lines of Ti II are concentrated at blue-optical to UV wavelengths \citep[e.g.][see their Figure C.1]{Hoeijmakers2019}, which MAROON-X does not cover. However, we note that given the observed presence of Fe II and the fact that Ti ionizes more readily than Fe, we expect Ti II to be present on \pla.

Table \ref{CCF results} provides a summary of the chemical species with clear or tentative observed signals. For each element or ion, we report confidence intervals derived from slices in $K_p$ and $V_\mathrm{sys}$ going through the position of the maximum SNR of the map. The values then correspond to the position at  $\mathrm{SNR = SNR_{max} - 1}$. The $K_p$ value of H I is notably missing from Table \ref{CCF results}, with its peak outside of our grid. Therefore, for H I only, we instead take the $V_\mathrm{sys}$ slice that correspond to $K_p = 187\,\mathrm{km\,s^{-1}}$, the expected $K_p$ of the planet, to calculate the uncertainties in $V_\mathrm{sys}$. 

\begin{figure*}[t]
  \centering
  \includegraphics[width=\textwidth]{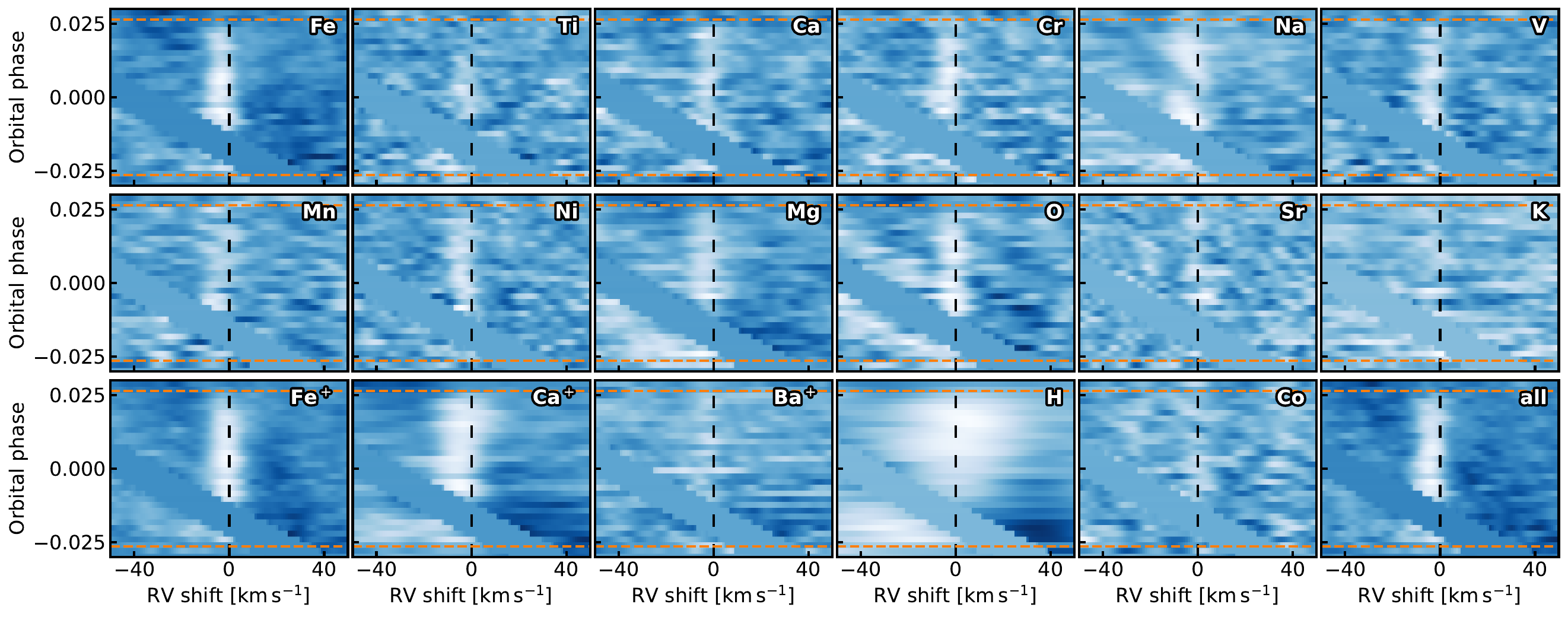}
  \caption{\textbf{Similar as Figure~\ref{RM mask}, but now in the planetary rest frame and showing individual species.}
  Orange horizontal dashed lines mark the beginning and the end of the transit event. The black dashed line indicates the expected position of the planet assuming a static atmosphere and no three-dimensional effects. The white trace corresponds to the signal from the planetary atmosphere. Most species show signals that are stable in both strength and blueshift throughout the transit duration of \pla.}
  \label{trails}
\end{figure*}

We observe a net blueshift of a few $\mathrm{km\,s^{-1}}$ in $V_\mathrm{sys}$ for all observed signals (Figure~\ref{detections}; Table \ref{CCF results}), consistent with the presence of day-to-night winds in the atmosphere of \pla\ likely driven by a large temperature gradient between both hemispheres. This is common for inflated planets on close-in orbits \citep{Showman2012, Wardenier2021, Wardenier2023, Gandhi2023, Akin2025}.  
Viewed in the rest frame of \pla, all species (with the exception of Sr I and H I) show signals that are systematically blueshifted (Figure \ref{trails}). Moreover, the planetary traces show near-constant strengths throughout the duration of the transits, with a slight decrease towards egress likely induced by the PCA subtraction. This is in contrast to slightly colder planets such as WASP-76b and WASP-121b which show a signals that become progressively stronger and more blueshifted during transit~\citep{Ehrenreich2020, kesseli_confirmation_2021, Pelletier2023, borsa_atmospheric_2021}. In this case, the more constant signal on \pla\ could indicate a more symmetric morning and evening terminator.  This could potentially be due to its overall hotter temperature reducing the vaporization timescale of any clouds circulated from the nightside to the morning terminator~\citep{savel_no_2022}, or increasing ionization rates, causing stronger magnetic drag and hence hampering the advection of heat away from the substellar point~\citep{Wardenier2021}.

We measure an elevated $K_p$ with respect to the expected $K_p = 187 \pm 4\,\mathrm{km\,s^{-1}}$ for some species, the most obvious being Ca II and H I (Figure \ref{detections}; Table \ref{CCF results}). This is not entirely surprising. Signals from different species can probe various parts of the atmosphere \pla. For instance, species with very strong lines can probe the extended, low-density upper atmosphere, while neutral species tend to probe deeper layers of atmosphere where they are more abundant~\citep[e.g.,][]{Maguire2022, Kesseli2022, Prinoth2023}. In addition, $K_p$ is intrinsically difficult to constrain from transit datasets owing to the relatively small portion ($\sim 7\%$) of the orbit sampled during transit, which leads to large uncertainties. Therefore, although some species appear to have an elevated $K_p$, the values remain consistent with the predicted orbital velocity within $1\sigma$. A similar behavior has also been observed on other UHJs, where values of $K_p$ can span a range of $\sim 60\,\mathrm{km\,s^{-1}}$ for different chemical species that probe distinct regions of the atmosphere \citep[e.g.,][see their Figure 6]{Borsato2023}.


\subsection{Free Retrieval Results}

There are many physical processes that can alter the line depths of observed chemical species, including the surface gravity and mean molecular weight via the scale height, the continuum level, the abundances, and the temperature-pressure profile. This creates degeneracies between fitted parameters in atmospheric retrievals that can be hard to break~\citep{Benneke2012}. In HRCCS, these degeneracies are aggravated by the data reduction process, which removes the absolute continuum level through normalization and detrending. As a result, the observed signal is mostly sensitive to relative line contrasts rather than absolute line depths \citep{Brogi2019, Gibson2020}. Therefore, we find that relative abundances are much better constrained, consistent with previous studies \citep{gibson_relative_2022, Maguire2022, Gandhi2023, Pelletier2023}. 
As a result, we present our results and derive all of our conclusions based on measured relative metal abundances (e.g., Mg/Fe) rather than `absolute' (e.g., Fe/H, Mg/H) abundances.

We follow the approach of \citet{Gandhi2023} to interpret our retrieval results. We derive the relative abundances of each species normalized to the solar value \citep{Asplund2009} using

\begin{equation}
\label{logabun}
    \varphi_\chi = \mathrm{log}(\chi/\chi_{\mathrm{Fe}}) - \mathrm{log}(\chi/\chi_{\mathrm{Fe}})_\odot,
\end{equation}

where $\chi$ is the abundance of the species, $\chi_{\mathrm{Fe}}$ is the abundance of Fe, and the subscript $\odot$ indicates solar values.  For $\varphi_\mathrm{Ti}$ and $\varphi_\mathrm{V}$, we include the sum of the contributions from both atomic and molecular species
(i.e. $\varphi_\mathrm{Ti} = \mathrm{log}(\chi_\mathrm{Ti \, I}+\chi_\mathrm{TiO})/\chi_\mathrm{Fe}) - \mathrm{log}(\chi_\mathrm{Ti \, I}+\chi_\mathrm{TiO})/\chi_\mathrm{Fe})_\odot$). 
We use the composition of the Sun as a reference as no stellar abundances are available in the literature for most species detected here. This is due to the rapid rotation of HAT-P-70, which causes severe spectral line broadening that hinders precise abundance measurements~\citep[e.g.][]{lam_secrets_2024}.

\begin{figure*}[t]
  \centering
  \includegraphics[width=\textwidth]{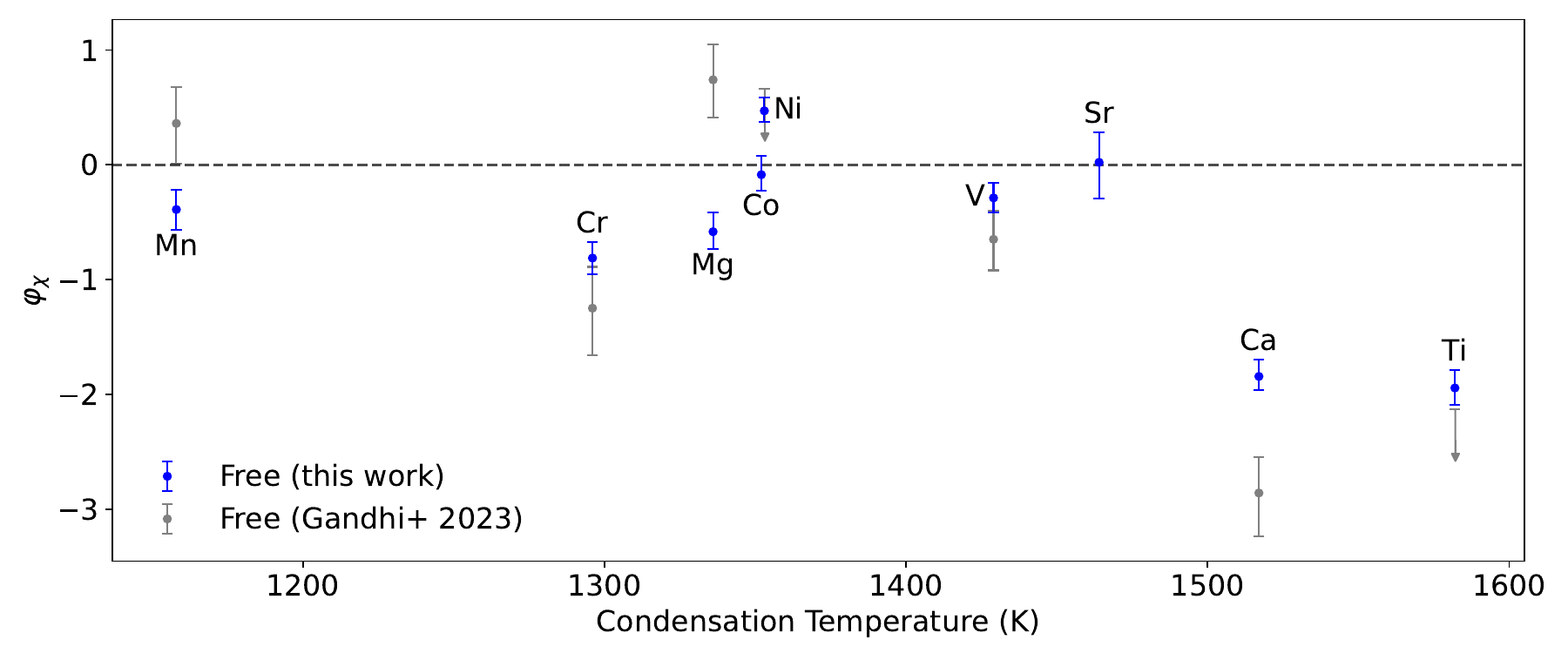}
  \caption{\textbf{Retrieved elemental abundances of metal species of interest using a free retrieval prescription.} Abundances are normalized according to Equation \ref{logabun} ($\varphi_\chi = \mathrm{log}(\chi/\chi_{\mathrm{Fe}}) - \mathrm{log}(\chi/\chi_{\mathrm{Fe}})_\odot$). Results from \citet{Gandhi2023} are shown in grey. All errorbars represent 1$\sigma$ uncertainties and arrows represent the retrieved 2$\sigma$ upper limits. We provide new constraints for Co and Sr, which were not considered in \citet{Gandhi2023}. Our value for Ti (V) is calculated by summing the retrieved abundances for atomic Ti (V) and molecular TiO (VO). Both works are consistent within the $2\sigma$ level.}
  \label{freeg23}
\end{figure*}

We use a free retrieval approach to constrain the abundances of nine metal species (Ti, Ca, Cr, Ni, Mg, V, Mn, Co, and Sr) relative to Fe and compare these with the results of \citet{Gandhi2023} (Figure \ref{freeg23}). 
Our results are overall consistent for most elements, although do show some notable discrepancies.
For $\varphi_\mathrm{Ti}$, our value sits slightly above the $2\sigma$ upper limit reported by \citet{Gandhi2023} (likely consistent at $3\sigma$), which included contributions from Ti, TiO and TiH. For $\varphi_\mathrm{Ni}$, we recover a slightly supersolar abundance, consistent with the reported upper limit from \citet{Gandhi2023}. For $\varphi_\mathrm{Ca}$, $\varphi_\mathrm{Cr}$ and $\varphi_\mathrm{V}$, we retrieve values slightly higher than those previously reported, but still consistent within $1-2\sigma$. In contrast, our retrieval favors subsolar values for $\varphi_\mathrm{Mg}$ and $\varphi_\mathrm{Mn}$, whereas \citet{Gandhi2023} reported solar-to-supersolar abundances. Finally, we report solar abundances for $\varphi_\mathrm{Co}$ and $\varphi_\mathrm{Sr}$, which were not considered in any previous analysis. Overall, both studies show broad consistency, with all values in agreement at roughly the $2\sigma$ level.

\subsection{Chemical Equilibrium Retrieval Results}

\begin{figure*}[t]
  \centering
  \includegraphics[width=\textwidth]{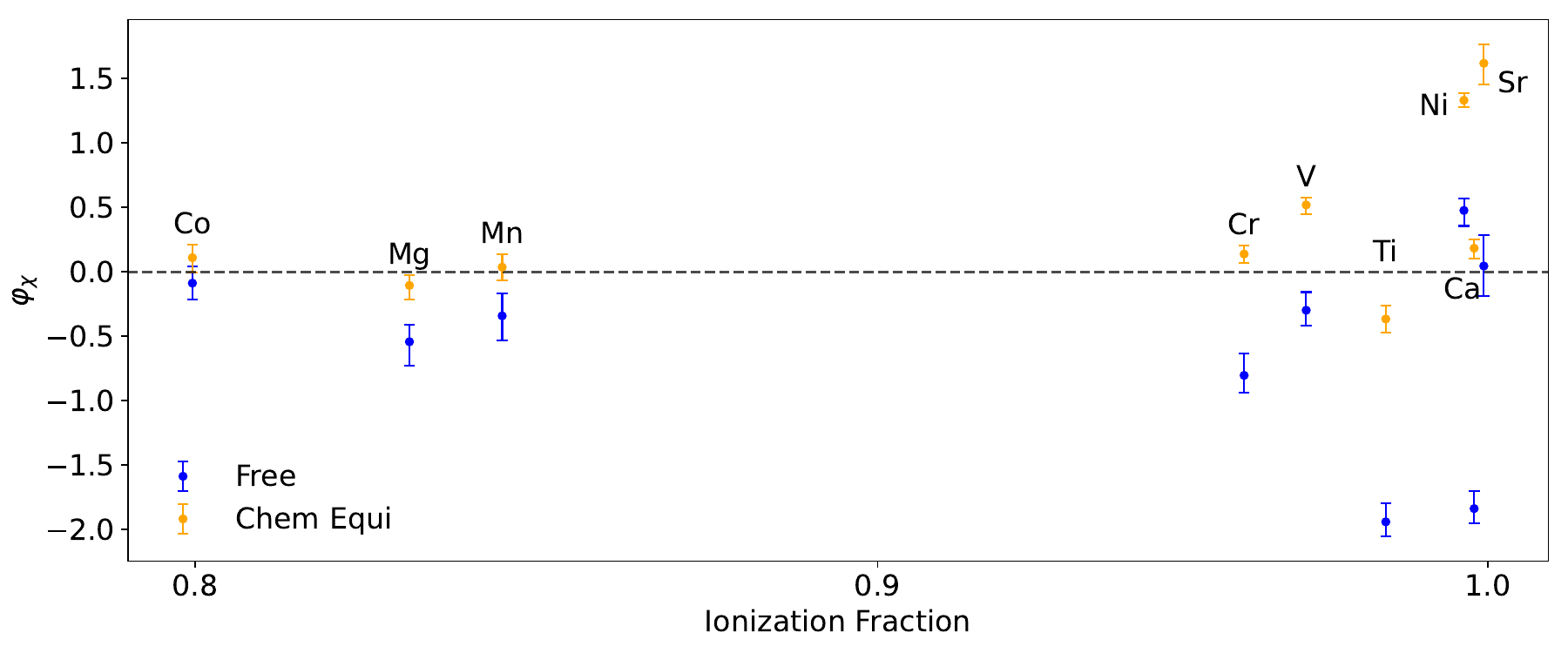}
  \caption{\textbf{Comparison of the retrieved elemental abundances with the free and chemical equilibrium prescriptions.} Abundances are normalized according to Equation \ref{logabun} ($\varphi_\chi = \mathrm{log}(\chi/\chi_{\mathrm{Fe}}) - \mathrm{log}(\chi/\chi_{\mathrm{Fe}})_\odot$). Species are ordered according to the predicted ionization fractions at $4000\,\mathrm{K}$ and $1\,\mathrm{mbar}$ by \texttt{FastChem} \citep{Stock2018, Stock2022}. All errorbars represent 1$\sigma$ uncertainties. In the free retrieval, Ti (V) is the sum of atomic Ti (V) and molecular TiO (VO). With an ionization fraction of 0.67, Fe is the least readily ionized species included in our retrievals. As a result, the free retrieval finds generally lower abundances relative to Fe than the equilibrium chemistry retrieval which accounts for ionization effects.}
  \label{freechem}
\end{figure*}

Free retrievals assuming constant-with-altitude abundance profiles can struggle to accurately infer abundances of chemical species that have highly non-uniform abundance profiles, for example due to ionization or thermal dissociation \citep[e.g.][]{brogi_roasting_2023}. In the context of UHJs, the upper atmosphere can heat up to several thousand Kelvin, where atoms may be strongly ionized (Figure \ref{corner chem}, top right panel). Of the species we include in our retrieval analyses, Fe has the highest ionization threshold. For instance, \texttt{FastChem} predictions at $4000\ \mathrm{K}$ and $1\ \mathrm{mbar}$ yield an ionization fraction of 0.67 for Fe and 0.98 for Ti (Figure \ref{freechem}). Therefore, abundances inferred from free retrievals assuming constant-with-altitude profiles should be interpreted with caution, especially for species expected to be strongly affected by ionization or dissociation. In such scenarios, it is more informative to compare retrieved abundances with predictions from equilibrium chemistry \citep{Gandhi2023, Chachan2025}. In order to incorporate the effects of ionization in our abundance estimation, while also allowing for non-solar metallic ratios, we also run chemical equilibrium retrievals as described in Section \ref{chemequi description}. While inferred abundances from free and chemical equilibrium retrievals should yield similar values for species that do not ionize readily such as Fe, Co, and Mg, elements with weakly bound outer electrons such as Ca, Sr, and Ti should be more affected by allowing for highly non-uniform ionized abundance profiles.

Indeed, we find that the change in retrieved abundances relative to Fe becomes increasingly significant for elements with higher ionization fractions (Figure \ref{freechem}). Here we define the ionization fraction of an element to be the ratio of the amount of singly ionized atoms to the total amount of atoms of that element at a given temperature and pressure. In this case, we calculate them using \texttt{FastChem} at a temperature of $4000\ \mathrm{K}$ and a pressure of $1\ \mathrm{mbar}$. For Co, Mg, and Mn (and Fe), which have high ionization thresholds, this correction is very small, and both the free retrieval assuming constant-with-altitude abundance profiles and the chemical equilibrium retrieval accounting for ionization find similar values. 
On the other hand, including the effect of thermal ionization provides a large correction in inferred abundances for chemical species with lower ionization potentials such as Ca and Sr. Indeed for most species (e.g., Mg, Mn, Cr, Ti, Ca), the inclusion of thermal ionization in the modeling naturally explains inferred underabundances from the free retrievals (which, again, should be thought of as lower limits), with values more consistent with abundance ratios of the solar photosphere~\citep{Asplund2009}. However, in the cases of V, Ni, and Sr, accounting for thermal ionization in an equilibrium chemistry framework appears to provide an over-correction with abundance ratios further away from the expected solar value.

Recently, \citep{Guo2025} also explored the atmospheric composition of \pla\ using free and equilibrium chemistry retrievals. However, in their case, relative abundance constraints from both retrieval prescriptions were consistent, with only minor corrections from accounting for ionized abundance profiles. This is likely because they observed the dayside thermal emission of \pla, which is sensitive to deeper layers of the atmosphere relative to transmission spectroscopy, where ionization is not as significant. In comparison, our transit data are sensitive to lower pressure regions of the atmosphere with hotter temperatures where atoms are expected to be ionized to a much higher degree. This demonstrates that biases that may arise from applying well-mixed retrievals to ultra-hot Jupiter atmospheres are not as severe in emission as in transmission spectroscopy.



\subsection{Supersolar V, Ni and Sr} \label{supersolar}

While the chemical equilibrium retrieval generally recovers relative abundances closer to solar values in the case of \pla, V, Ni, and Sr are notable exceptions (Figure \ref{freechem}, Table \ref{retrieval values}). Using a free retrieval, we find a slightly subsolar $\varphi_\mathrm{V} = -0.30^{+0.14}_{-0.12}$, a supersolar $\varphi_{\mathrm{Ni}} = 0.48^{+0.09}_{-0.12}$, and a solar $\varphi_{\mathrm{Sr}} = 0.04^{+0.24}_{-0.24}$. The chemical equilibrium retrieval accounting for ionization boosts these values to $\varphi_\mathrm{V} = 0.52^{+0.06}_{-0.07}$, $\varphi_\mathrm{Ni} = 1.33^{+0.05}_{-0.05}$, and $\varphi_\mathrm{Sr} = 1.62^{+0.15}_{-0.16}$, respectively. 
For vanadium, one possible explanation could be VO biasing the abundance inferences owing to its known line list imperfections~\citep{regt_quantitative_2022}.  Interestingly, a V enhancement was also reported in thermal emission observations of the UHJ WASP-121b using a similar chemical equilibrium retrieval approach and including VO \citep{Smith2024}. In that work, the authors postulated that the VO line list being incomplete could result in the total amount of V atoms in the atmosphere being overestimated. 
Indeed, if we calculate $\varphi_\mathrm{V}$ considering only V~I (not including VO), we find an enrichment level that is more consistent with expectations from a solar-like composition.

The discrepancy in the recovered $\varphi_\mathrm{Ni}$ and $\varphi_\mathrm{Sr}$ abundances is both more extreme and more puzzling. Interestingly, \citet{Guo2025} also retrieve a high $\varphi_\mathrm{Ni} = 1.01^{+0.50}_{-0.60}$ on the dayside of \pla\ using a chemical equilibrium prescription, which the authors attribute to the potential accretion of Ni-rich planetesimals.  
While this could explain a slight enrichment of Ni, such a hypothesis cannot explain Ni being overabundant relative Fe by an order of magnitude or more~\citep{Pelletier2023}. Instead, our measured elevated $\varphi_\mathrm{Ni}$ is more likely a result of atmospheric processes or model shortcomings.
We therefore explore potential biases that could arise due to aliasing and model incompleteness. 
Although Ni and Sr have many spectral lines in the MAROON-X bandpass, the observed signals are likely only from a few of the stronger lines that lie above the continuum level set by H$^{-}$ bound-free absorption in modeled transmission spectra. It is then possible that the atmospheric inference is driven primarily by relatively few spectral features in comparison to other species such as Fe which have a larger number of strong lines. 
In this case, aliasing effects, where lines coincidentally overlap in wavelength with the lines of other chemical species not included in our model could more easily bias our results~\citep[e.g.][]{Borsato2023}. Having only a few observable spectral features also increases the risk of biases from possible outliers or residuals remaining in the data. It is therefore possible that the VMR of Ni and Sr relative to Fe are artificially increased in the retrieval as a result of a few lines being contaminated.

To explore potential sources of aliasing, we search through the cross sections of all available species in the DACE database\footnote{\url{https://dace.unige.ch/opacityDatabase/}} and identify elements with strong lines overlapping with those of Ni or Sr. For Ni, we find that Co I and Ce II (ionized cerium) have some notable features that overlap with the strongest Ni lines. While Co I already included in our retrievals, we choose to not further include Ce II given that our retrievals struggle with ionic species. Similarly for Sr, we identify Co I and Mn I as species with the most overlapping lines, both of which are included in our retrievals. Nevertheless, we cannot rule out other unknown absorbers as sources of contamination. To mitigate possible contamination from Ce II and to further guard against aliasing from coincidental alignment with any potential unknown spectral features not included in our models, we repeat our analysis but with the wavelengths of the five strongest absorption lines of Ni I and Sr I in the MAROON-X bandpass masked out from the data. The width of the mask centered on each line corresponds to a $\pm 200\,\mathrm{km\,s^{-1}}$ RV shift. In both cases, the values of $\varphi_\mathrm{Ni}$ and $\varphi_\mathrm{Sr}$ remain elevated despite omitting the deepest lines from the analysis, suggesting that the high inferred abundance ratios are not the result of only a few absorption features driving the signal.  We also tested masking additional strong lines of Ni, finding that $\varphi_{\mathrm{Ni}}$ remains consistently supersolar in chemical equilibrium retrievals, albeit with larger error bars. For \pla, both transmission (this work) and emission \citep{Guo2025} spectroscopy recover an elevated Ni abundance. Interestingly, enrichments of Ni relative to Fe are commonly inferred from transmission spectroscopy atmospheric retrievals of UHJs~\citep{hoeijmakers_hot_2020, Gandhi2023, Pelletier2023} and generally show strong signals in emission spectroscopy~\citep{kasper_unifying_2022, hoeijmakers2024, pelletier_crires_2025}. 




Rather than thermal ionization, which is accounted for in the equilibrium chemistry retrievals, photoionization could provide a partial explanation to the observed elevated Ni abundance.  With its higher overall opacity in the UV relative to Ni I, Fe I may be more susceptible to photoionization by high-energy stellar irradiation. As this effect is not considered in our equilibrium chemistry models, the inferred amount of Ni relative to Fe could, in turn, appear to be enhanced with respect to expectations. However, it is unclear that photoionization alone could explain the observed $\sim 20\times$ enhancement of Ni we measure.  Additionally, this would further assume that photoionization affects other metals (Co, Mg, Mn, etc.) similarly in order to preserve their inferred solar proportions relative to Fe. Other possibilities include either disequilibrium chemical processes in \pla's atmosphere or incorrect modeling of the Ni chemistry.

\begin{table}[t]
\centering
\renewcommand{\arraystretch}{1.1}
\setlength{\tabcolsep}{15pt}
\begin{tabular}{ccc}
\hline
\hline
Parameter & Free & Chem Equi\\
\hline
$\varphi_\mathrm{Ti}$ & $-1.94^{+0.15}_{-0.11}$ & $-0.37^{+0.10}_{-0.10}$\\
$\varphi_\mathrm{Ca}$ & $-1.84^{+0.13}_{-0.12}$ & $0.18^{+0.07}_{-0.08}$\\
$\varphi_\mathrm{Cr}$ & $-0.80^{+0.17}_{-0.13}$ & $0.14^{+0.07}_{-0.07}$\\
$\varphi_\mathrm{Ni}$ & $0.48^{+0.09}_{-0.12}$ & $1.33^{+0.05}_{-0.05}$\\
$\varphi_\mathrm{Mg}$ & $-0.54^{+0.13}_{-0.19}$ & $-0.11^{+0.08}_{-0.11}$\\
$\varphi_\mathrm{V}$ & $-0.30^{+0.14}_{-0.12}$ & $0.52^{+0.06}_{-0.07}$\\
$\varphi_\mathrm{Mn}$ & $-0.34^{+0.17}_{-0.19}$ & $0.04^{+0.08}_{-0.10}$\\
$\varphi_\mathrm{Co}$ & $-0.09^{+0.13}_{-0.12}$ & $0.11^{+0.10}_{-0.12}$\\
$\varphi_\mathrm{Sr}$ & $0.04^{+0.24}_{-0.24}$ & $1.62^{+0.15}_{-0.16}$\\
$K_p/\mathrm{km\,s^{-1}}$ & $186.36^{+0.74}_{-0.77}$ & $186.04^{+0.98}_{-0.92}$\\
$V_{\mathrm{sys}}/\mathrm{km\,s^{-1}}$ & $18.48^{+0.07}_{-0.05}$ & $18.50^{+0.09}_{-0.07}$\\
$\mathrm{M_p}/\mathrm{M_{Jup}}$ & - & $1.64^{+0.28}_{-0.27}$\\
$v_\mathrm{rot}/\mathrm{km\,s^{-1}}$ & $4.44^{+0.23}_{-0.19}$ & $4.34^{+0.28}_{-0.19}$\\
\hline
\hline
\end{tabular}
\caption{\textbf{Summary of retrieved parameters with $1\sigma$ uncertainties.} Values of $\varphi_\chi$ are derived parameters from the retrievals.}
\label{retrieval values}
\end{table} 


We also explore the possibility that the observed enrichment in Ni could be due to opacity inaccuracies. While our main analysis uses the VALD \citep{piskunov_vald:_1995, ryabchikova_major_2015} database for atomic cross sections, we verify our results by repeating our analysis using cross sections from the Kurucz \citep{kurucz_table_1975, kurucz_including_2017} and NIST \citep{kramida_atomic_2009} databases. Similarly to \citet[][see their Extended Data Fig.~7]{Pelletier2023}, we find no significant changes in the results and recover an elevated $\varphi_\mathrm{Ni}$ in all cases.

\subsection{Partial Nightside Cold-Trapping?}

\begin{figure}[t]
  \centering
  \includegraphics[width=0.5\textwidth]{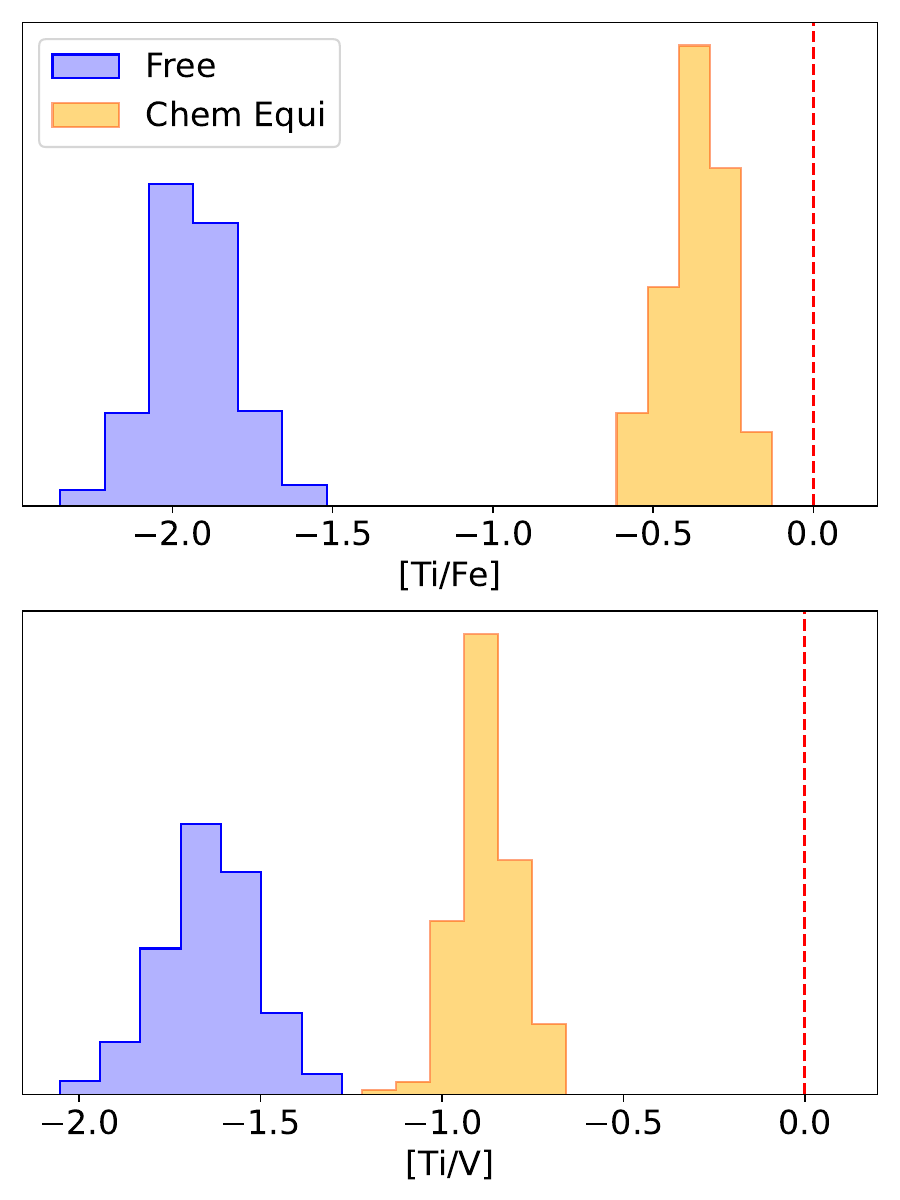}
  \caption{\textbf{Retrieved [Ti/Fe] (top) and [Ti/V] (bottom) abundance ratios from both retrieval prescriptions}. In the free retrieval, Ti (V) is the sum of atomic Ti (V) and molecular TiO (VO).  In the chemical equilibrium retrieval, a full chemical network of Ti, Fe, and V species is considered. The red vertical dashed line indicates the  Ti/Fe and Ti/V elemental ratios in the solar photosphere.  As Ti ionizes more readily, it appears highly underabundant relative to Fe in the free retrieval where non-uniform ionized abundance profiles cannot be reproduced. Accounting for ionization in the chemical equilibrium retrieval can explain part of the underabundance, but with [Ti/Fe] remaining slightly subsolar.  Comparing Ti to more chemically similar V, [Ti/V] is more significantly underabundant, although this may also be due to V being enriched.}
  \label{Ti/V}
\end{figure}

Past studies have shown that abundances of highly refractory chemical species could be underabundant, or even depleted, if the nightside temperature of the UHJ is low enough for condensation and rain-out to occur~\citep{parmentier_3d_2013, Ehrenreich2020}. One notable example is titanium, which is often missing from the observed chemical inventory~\citep{Merritt2020, Kesseli2022}, or measured to be underabundant or depleted~\citep[e.g.,][]{Gandhi2023, Pelletier2023, pelletier_enriched_2026, Prinoth2025} in UHJs with equilibrium temperatures below $\sim$ 2400\,K. Meanwhile, planets with T$_{\mathrm{eq}}$ $>$ 2600\,K tend to show strong signatures of Ti, Ti$^+$, and TiO in their atmospheres~\citep{Prinoth2022, Prinoth2023, nugroho_high-resolution_2017, Hoeijmakers2018, cont_detection_2021}, suggesting that their nightsides are too hot for even ultra-refractory Ti to condense. \pla\ notably sits in this transition region at $\mathrm{T_{eq} = 2562\,K}$ where part of its titanium may still be cold-trapped despite Ti-species being observed in its transmission spectrum~\citep{BelloArufe2022}. Using a free retrieval, we find the Ti/Fe in the atmosphere of \pla\ to be below the Ti/Fe of the Sun by $1.94^{+0.11}_{-0.15}$ dex, consistent with the upper limit found by \citet{Gandhi2023}. However, our chemical equilibrium retrieval refines this to a slightly subsolar value at $-0.37^{+0.10}_{-0.10}$ dex when accounting for contributions from all titanium-bearing species in the chemical network, including Ti$^{+}$, TiO, TiO$_2$, and TiH (Figure \ref{Ti/V}, Top panel). Indeed, accounting for thermal ionization under the assumption of chemical equilibrium, $\varphi_\mathrm{Ti}$ is only marginally subsolar. Because Fe ionizes less readily than Ti, the free retrieval can underestimate the proportion of titanium relative to iron in highly ionized ultra-hot Jupiter atmospheres. Although Ti/Fe is much closer to the solar ratio under chemical equilibrium assumptions, it remains mildly subsolar, suggesting that ionization effects alone may not fully account for the inferred low Ti abundance.

As a secondary check, we also compare Ti to V, a chemically similar species with comparable ionization potential that forms analogous oxides (TiO, TiO$_2$, VO, VO$_2$). Our free retrieval finds Ti/V depleted by a factor of $1.65^{+0.16}_{-0.17}$ dex compared to the solar ratio, and our chemical equilibrium retrieval finds Ti underabundant by $0.88^{+0.08}_{-0.09}$ dex (Figure \ref{Ti/V}, Bottom panel).
The value in the free retrieval is derived from the sum of the VMRs of Ti (V) and TiO (VO) to have a better representation of the entire elemental content rather than just the neutral atomic species.
However, we caution that the inferred Ti/V relative abundance may be affected by uncertainties in the TiO and VO line lists (see Section \ref{supersolar} and \citealp{Prinoth2022}). We note that similarly low Ti/V ratios have been reported in other UHJs, even based on space-based low resolution observations which should not be as affected by line list issues (e.g. WASP-121b; \citealp{pelletier_enriched_2026}).

Meanwhile, $\mathrm{Ti^+}$ was detected by cross-correlation in \pla's atmosphere \citep{BelloArufe2022}. Combined with our observed signal of neutral Ti, we can rule-out titanium being completely depleted by cold-trapping. If any rainout of titanium does occur on the nightside of \pla, the depletion is only partially effective, with mixing and advection timescales being efficient enough to circulate and evaporate titanium back to the gas phase in the observable photosphere of the terminator regions.

\subsection{Constraints on Planetary Mass}

Currently, \pla\ only has an upper limit on the known mass \citep{Zhou2019}. It has been shown that the planetary mass can be constrained from atmospheric retrievals via constraints on the scale height, although it is degenerate with multiple fitted parameters \citep{Gandhi2023}. In this work, we constrain the planetary mass using our chemical equilibrium retrieval, where $\mathrm{M_p}$ is treated as a free parameter. Under this framework, our retrieval recovers a bimodal distribution, with two solutions centered around $\sim 1.4\,\mathrm{M_{Jup}}$ and $\sim 1.8\,\mathrm{M_{Jup}}$, both of which are comparably explored by the sampler (Figure \ref{corner chem}). The marginalized constraint of $\mathrm{M_p = 1.64^{+0.28}_{-0.27}\,M_{Jup}}$ therefore reflects a combination of both modes instead of a single well-defined solution. We note that both modes are consistent with the spectroscopically derived value of $\mathrm{M_p = 1.66^{+0.20}_{-0.20}\,M_{Jup}}$ reported by \citet{Gandhi2023}, as well as the known upper limit ($\mathrm{M_p} < 6.78\ \mathrm{M_{Jup}}$).

The presence of a bimodal solution indicates that the data do not uniquely constrain the planetary mass under our modeling assumptions. In particular, this behaviour arises from the degeneracies between planetary mass, temperature structure, and chemical abundances through their combined effect on the atmospheric scale height and line contrast. A lower-mass solution can be compensated by a cooler temperature structure or reduced abundances. Consistent with this picture, the posterior distributions of Fe and several other species exhibit a secondary peak, tracking the bimodality in $\mathrm{M_p}$. Overall, our results further highlight that spectroscopically inferred planetary masses remain sensitive to degeneracies with the temperature structure and chemical composition of the atmosphere.

\section{Conclusions} \label{conclusions}

In this paper, we present a detailed atmospheric analysis of the ultra-hot Jupiter \pla\ based on two transit observations with the high-resolution red-optical MAROON-X spectrograph. Using cross-correlation spectroscopy, we detect a rich set of atomic species in the planet's atmosphere. Our analysis reveals clear signals from 14 neutral and ionized species, including Fe I, Fe II, Ti I, Ca I, Ca II, Cr I, Na I, V I, Mn I, Ni I, Mg I, Ba II, O I, and Sr I, while also providing tentative evidence for H I, Co I, and K I.

To quantify the atmospheric composition, we perform atmospheric retrievals at high spectral resolution and constrain the relative abundances of nine metal species. We explore both free and chemical equilibrium retrieval frameworks in order to assess the impact of modeling assumptions on the inferred abundances. We find that the free retrieval, which assumes constant-with-altitude abundance profiles, tends to underestimate the abundances of species with relatively low ionization potentials compared to Fe. In contrast, the chemical equilibrium retrieval naturally accounts for thermal ionization and yields relative abundances that are generally closer to solar ratio. Notable exceptions are Ni/Fe and Sr/Fe, for which the chemical equilibrium retrieval consistently finds enrichment factors of $\sim 20$ and $\sim 40$ relative to solar, respectively, potentially indicating the influence of disequilibrium processes in \pla's atmosphere.

Titanium provides a useful probe for condensation and ionization processes. Both Ti and Ti$^+$ have been detected in the atmosphere, ruling out complete depletion by nightside cold trapping. However, even with thermal ionization accounted for, the inferred abundance ratios remain subsolar (Ti/Fe $\sim -0.4$ dex and Ti/V $\sim -0.9$ dex), suggesting partial depletion. Given that \pla\ lies in the transition regime for titanium condensation, this result is consistent with partial nightside cold trapping combined with efficient atmospheric mixing that replenishes Ti in the observable terminator.

In addition, our chemical equilibrium retrieval provides constraints on the planetary mass. The posterior distribution is bimodal, with two solutions near $\sim 1.4\,\mathrm{M_{Jup}}$ and $\sim 1.8\,\mathrm{M_{Jup}}$. The marginalized constraint of $\mathrm{M_p = 1.64^{+0.28}_{-0.27}\,M_{Jup}}$ reflects the combination of these two modes and remains consistent with both the previously reported spectroscopic estimate and the current upper limit on the planetary mass. This result illustrates the potential of high-resolution transmission spectroscopy to provide constraints on planetary bulk properties, while also highlighting the sensitivity of such measurements to degeneracies with the atmospheric temperature structure and composition.

Together, these results place \pla\ in the transition regime between cooler UHJs with potential titanium cold traps and hotter systems where titanium remains fully in the gas phase. The chemical inventory and abundance measurements presented here provide new insights into the atmospheric chemistry, dynamics, and structure of one of the hottest known Jupiters, and set the stage for future comparative studies of exoplanet atmospheres in extreme irradiation regimes.

\begin{acknowledgements}
Based on observations obtained at the international Gemini Observatory, a program of NSF NOIRLab, which is managed by the Association of Universities for Research in Astronomy (AURA) under a cooperative agreement with the U.S. National Science Foundation on behalf of the Gemini Observatory partnership: the U.S. National Science Foundation (United States), National Research Council (Canada), Agencia Nacional de Investigaci\'{o}n y Desarrollo (Chile), Ministerio de Ciencia, Tecnolog\'{i}a e Innovaci\'{o}n (Argentina), Minist\'{e}rio da Ci\^{e}ncia, Tecnologia, Inova\c{c}\~{o}es e Comunica\c{c}\~{o}es (Brazil), and Korea Astronomy and Space Science Institute (Republic of Korea). This work was enabled by observations made from the Gemini North telescope, located within the Maunakea Science Reserve and adjacent to the summit of Maunakea. We are grateful for the privilege of observing the Universe from a place that is unique in both its astronomical quality and its cultural significance. 
This project has been carried out within the framework of the National Centre of Competence in Research PlanetS supported by the Swiss National Science Foundation (SNSF) under grant 51NF40\_205606. The authors acknowledge the financial support of the SNSF. This work was partially funded by the French National Research Agency (ANR) project EXOWINDS (ANR-23-CE31-0001-01). The MAROON-X group acknowledges funding from the David and Lucile Packard Foundation, the Heising-Simons Foundation, the Gordon and Betty Moore Foundation, the Gemini Observatory, and the NSF (award numbers 2108465 and 2307177). J.P.W.\ acknowledges support from the Trottier Family Foundation via the Trottier Postdoctoral Fellowship, as well as support from the Canadian Space Agency (CSA) under grant 24JWGO3A-03.
This publication makes use of The Data \& Analysis Center for Exoplanets (DACE), which is a facility based at the University of Geneva (CH) dedicated to extrasolar planets data visualization, exchange and analysis. DACE is a platform of the Swiss NCCR PlanetS, federating the Swiss expertise in Exoplanet research. The DACE platform is available at \url{https://dace.unige.ch}.

\bigbreak
\textit{Software}: \texttt{NumPy} \citep{harris_array_2020}, \texttt{SciPy} \citep{virtanen_scipy_2020}, \texttt{Matplotlib} \citep{hunter_matplotlib_2007}, \texttt{Astropy} \citep{astropy_collaboration_astropy_2013, astropy_collaboration_astropy_2018}, \texttt{emcee} \citep{foreman-mackey_emcee_2013}, \texttt{corner} \citep{foreman-mackey_cornerpy_2016}

\end{acknowledgements}

\bibliography{sample7.bib, references.bib}{}
\bibliographystyle{aasjournalv7.bst}

\appendix
\setcounter{figure}{0}
\renewcommand{\thefigure}{A\arabic{figure}}
\setcounter{table}{0}
\renewcommand{\thetable}{A\arabic{table}}

\begin{figure*}[h]
  \centering
  \includegraphics[width=0.8\textwidth]{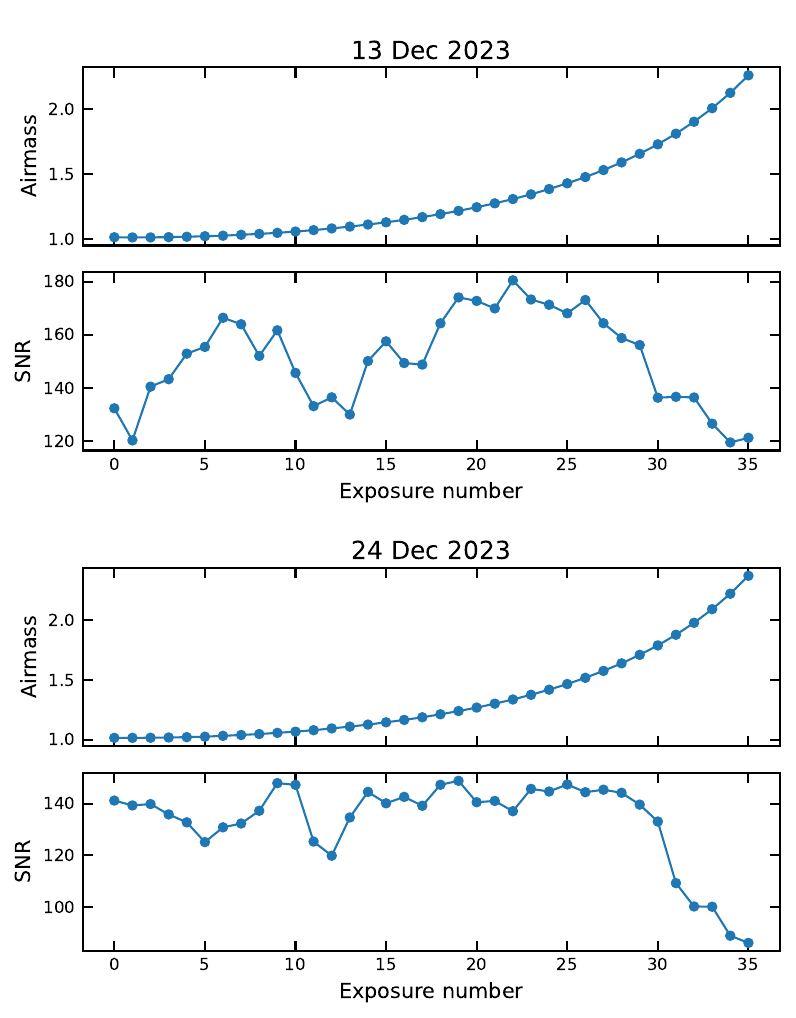}
  \caption{\textbf{Observation conditions for the nights of 13 Dec 2023 and 24 Dec 2023, including the airmass and SNR.} The reported SNR is mean SNR per order for a given exposure.}
  \label{conditions13}
\end{figure*}


\begin{figure*}[h]
  \centering
  \includegraphics[width=\textwidth]{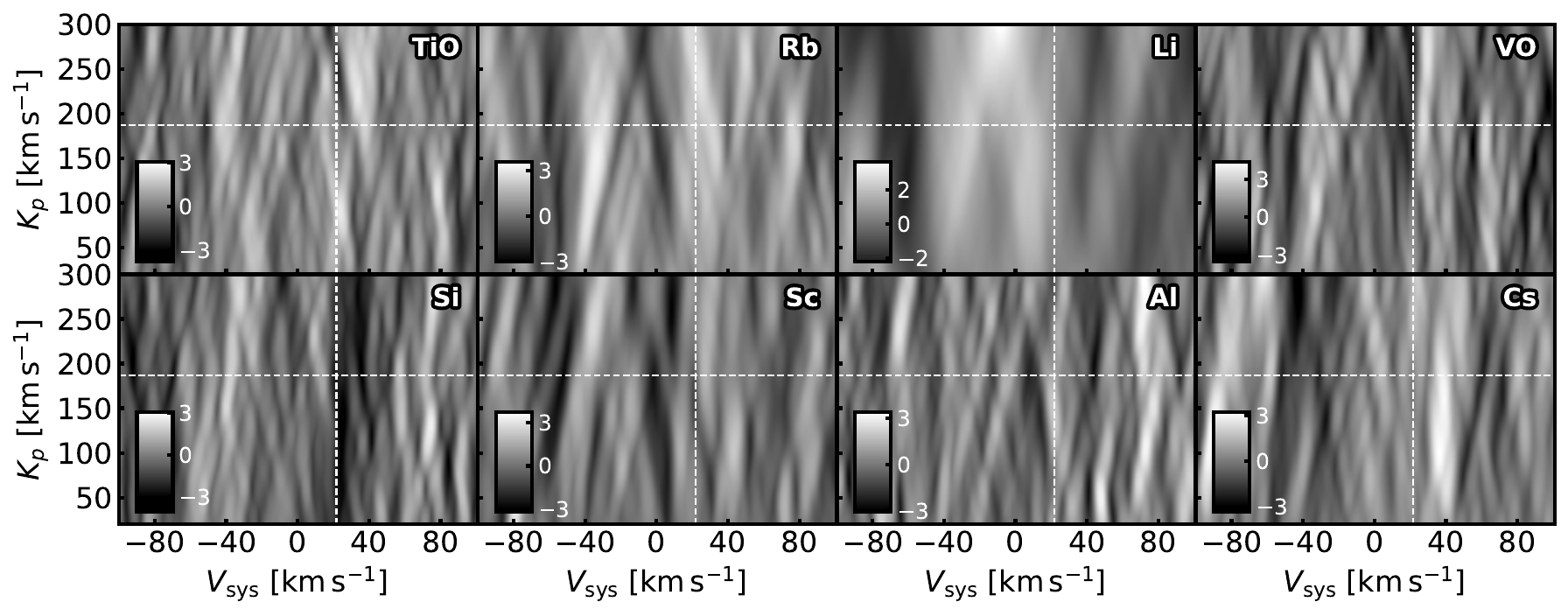}
  \caption{$\boldsymbol{K_p\ \textendash\ V_\mathrm{sys}}$\ \textbf{maps for species which we do not detect in \pla's atmosphere.}}
  \label{non-detections}
\end{figure*}

\begin{figure*}[h]
  \centering
  \includegraphics[width=\textwidth]{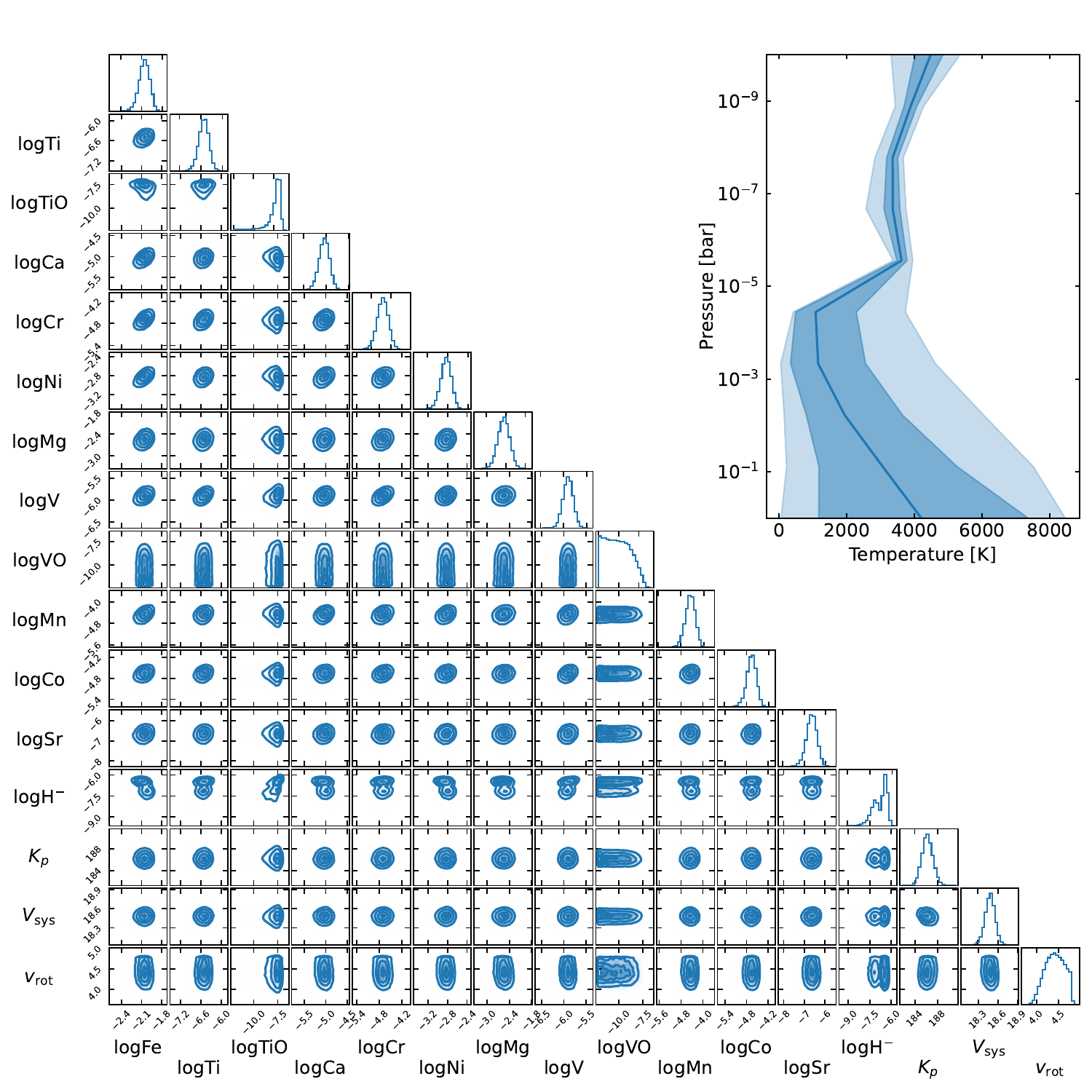}
  \caption{\textbf{Retrieved constraints for \pla's atmosphere and orbital properties with two MAROON-X transits using the free retrieval prescription.} \textbf{Lower left:} Corner plot including the marginalized distributions for the abundances of retrieved chemical species, radial velocity semi-amplitude $K_p$, systemic velocity $V_\mathrm{sys}$, and line broadening due to the planetary rotation $v_\mathrm{rot}$. \textbf{Top right:} Retrieved temperature structure with 8 points evenly distributed between $10^{-10}$ and $1$ bar. Shaded regions represent $1\sigma$ and $2\sigma$ confidence intervals.}
  \label{corner free}
\end{figure*}

\begin{figure*}[h]
  \centering
  \includegraphics[width=\textwidth]{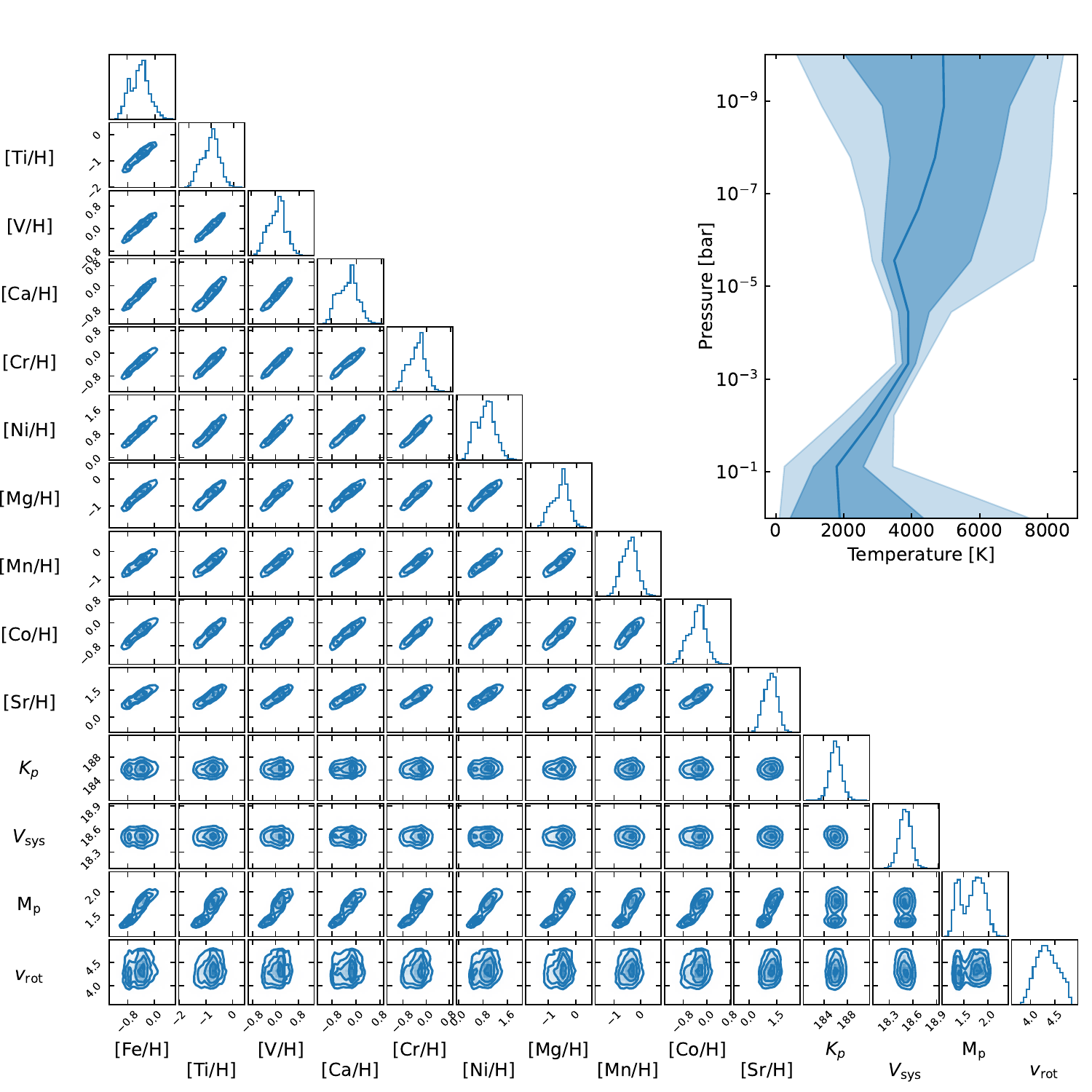}
  \caption{Same as Figure \ref{corner free}, but with the chemical equilibrium retrieval prescription and the planetary mass $\mathrm{M_p}$ as a free parameter.}
  \label{corner chem}
\end{figure*}



\end{document}